\documentclass[aps,amsmath,amssym,amsthm,superscriptaddress,reprint,secnumarabic]{revtex4-1}
\pdfoutput=1
\usepackage[colorlinks=true,
linkcolor=blue,
urlcolor=blue,
citecolor=blue]{hyperref}
\usepackage{cancel}
\usepackage{graphicx}
\usepackage{subcaption}
\usepackage{dcolumn}
\usepackage{bm}
\usepackage{url}
\usepackage{bm}   
\usepackage{bbm}   
\usepackage{verbatim}
\usepackage{stmaryrd}
\usepackage{amsthm}
\usepackage{amssymb}
\usepackage{array,multirow}
\usepackage{empheq}
\usepackage[usenames,dvipsnames]{color}
\usepackage[normalem]{ulem}
\usepackage{tikz}
\usepackage{enumitem}
\usepackage{algorithm}
\usepackage{algcompatible}
\usepackage{algpseudocode}

\renewcommand{\thesection}{\arabic{section}}
\renewcommand{\thesubsection}{\thesection.\arabic{subsection}}
\renewcommand{\thesubsubsection}{\thesubsection.\arabic{subsubsection}}
\makeatletter
\renewcommand{\p@subsection}{}
\renewcommand{\p@subsubsection}{}
\makeatother

\theoremstyle{plain}

\theoremstyle{definition}

\theoremstyle{remark}

\newcommand{\LS}{Szil\'{a}rd}

\begin{document}

\title{Partially Observable \LS\ Engines}

\author{Susanne Still}
\email{sstill@hawaii.edu}
\affiliation{Department of Physics and Astronomy}
\affiliation{Department of Information and Computer Sciences\\ University 
of Hawaii at M\=anoa, Honolulu, HI 96822, USA}

\author{Dorian Daimer}
\email{ddaimer@hawaii.edu}
\affiliation{Department of Physics and Astronomy}

\begin{abstract}
Almost a century ago, Leo \LS\ replaced Maxwell's ``demon" by machinery. The resulting information engine concept laid the foundation for studying the physical nature of information. \LS\ reduced the demon's role to mapping an observable onto a work extraction protocol, thereby creating and utilizing a meta-stable memory. While \LS\ demonstrated that this map can be implemented mechanistically, it was not chosen automatically by the demon, or ``observer", but rather given {\em a priori}. This choice implements the demon's intelligence. In \LS's original setup, the choice is trivial, but we show here that nontrivial data representations emerge for generalized, partially observable \LS\ engines. Partial observability is pervasive in real world systems with limited sensor types and information acquisition bandwidths. Generalized information engines may run work extraction at a higher temperature than memory formation, which enables the combined treatment of heat- and information engines. To date, \LS's (fully observable) information engine still serves as a canonical example. Implications of partial observability are under-explored, despite their ubiquitous nature. We provide here the first physical characterization of observer memories that result in minimal engine dissipation. We introduce a new canonical model, simple yet physically rich: a minor change to \LS's engine---inserting the divider at an angle---results in partially observable engines. We demonstrate how the demon's intelligence can be automated. For each angle and for each temperature ratio, an optimal memory is found algorithmically, enabling the engine to run with minimal dissipation. While naive coarse graining is sufficient for the special case of full observability, in general, minimally dissipative observers use {\em probabilistic} memories. We propose a simple model for an implementation of these memories, and construct a nontrivial physical codebook. We characterize the performance of engines with minimally dissipative memories, and compare their quality to that of engines using an optimized coarse graining of the observable. 
\end{abstract}

\maketitle

Information engines convert information to work and vice versa. They are useful theoretical constructs to address questions about the physical nature of information. Their origins go back to Maxwell's iconic thought experiment, often referred to as ``Maxwell's demon", which was conceived in the context of discussions about the role information plays in thermodynamics, and the statistical nature of the Second Law \cite{maxwell-demon,maxwell1871theory}. 

\LS\ contributed to this discussion a concrete physical mechanism in which the sentient being (the ``demon" that plays the role of observer) is replaced by machinery. \LS's idea can be summarized as follows: a divider is introduced into the center of a container with a one particle gas, reducing the accessible volume by a factor of two. The gas can then do work, on average up to $kT \ln(2)$ joules, where $k$ is the Boltzmann constant, in a process of isothermal, quasi-static expansion, at temperature $T$. The divider is used as a piston that moves into the empty half of the container in the process. 
For this to be possible, the observer needs to know which side of the container is empty. 
\LS\ pointed out that the observer's function is to (1) create a memory that is stable over the duration of work extraction, and (2) use this memory to make a decision, which exploits the expanding gas and makes the engine do work. This can be implemented by machinery, such as a switch, and pulleys appropriately attached to a weight---the engine needs no sentient being.

\LS\, and many others since, pointed out that it is the creation and destruction of the meta-stable memory, which adds to the entropy production of the overall process \cite{szilard, Landauer1961, demonbook90, bennett2003notes, Leff2003}.  
The second law is not circumvented by any such information engine, as long as the energy bill of the ``observer" mechanism is taken into account. This needs to be done, because the observer is an integral part of the engine, without which it could not function. 

Recent experimental efforts have succeeded in building devices which show that such information engines are feasible \cite{toyabe2010experimental, berut2012experimental, koski2014experimental, koski2014SEszilard, exp-landauer2014, koski2015chip, martinez2016brownian, hong2016experimental, camati2016experimental, gavrilov2016erasure, gavrilov2017direct, chida2017power, cottet2017observing, ciampini2017experimental, kumar2018nanoscale, paneru2018lossless, admon2018experimental, ribezzi2019large, peterson2020implementation, paneru2020efficiency, paneru2020colloidal, saha2020maximizing, dago2021information}, 
and that they can achieve the ultimate information-to-work conversion limit of $kT \ln(2)$ joules per bit of information, a limit dictated by the second law of thermodynamics \cite{szilard, Landauer1961, parrondo2015thermodynamics}. 

There is a sizable and rapidly increasing body of literature on information engines. Nonetheless, \LS's engine still serves as a canonical example.  
The ``observer" in these engines is reduced to machinery, but its ``intelligence" is implemented {\em a priori} by the designer. This intelligence manifests in the decision making, which is determined by the mapping from observable data to a meta-stable memory. The designer decides to choose the optimal map. But could that also be done automatically, by an algorithm?  

For the \LS\ engine, that optimal map from data to memory is simple: coarse grain the observed particle position into left/right of the divider. But imagine, for a moment, a different map (corresponding to a different ``observer"), namely coarse grain into 3 regions of equal size, two of which are on either side of the divider, and the third covers the center. On average, that observer would be able to extract less work, namely only ${2\over 3} kT \ln(2)$ joules. This shows that there are, in principle, ways in which an observer could err by choosing a suboptimal data representation. But this fact might get overlooked easily in the context of \LS's engine, because the most efficient data representation is trivial.

This is the case, because \LS\ assumed that the particle's position along the axis perpendicular to the divider is observable. That assumption implies that all of the information that is necessary to extract the maximal amount of average work is trivially available, because it is contained in the observable data. The ensuing literature followed in the same footsteps, and as a result, models of engines that we may call fully observable information engines assume that all the quantities which the observer needs to know to enable full work extraction are available (albeit potentially corrupted by noise). 

However, there exists a larger class of {\em partially observable} information engines, for which observables are correlated to some degree with those quantities. For this more general class of information engines \cite{CB}, the lowest possible entropy production is proportional to the amount of unuseable information retained in memory, {even in the limit of negligible measurement noise}. Fully observable information engines are a subclass, trivially able to retain zero useless information {in the noise free limit}, whereby they can reach the {ultimate} limit of zero entropy production in quasi-static operation. 

The need to make decisions based on {\em partial}  knowledge is ubiquitous in real world situations, where the choice of the observer's data representation matters. A central question, relevant to machine learning and decision making under uncertainty is: which strategy is the best? This question opens up deep problems, because one has to specify a metric that measures ``goodness", or, colloquially, a sense in which one observer is doing a better job than another observer. 

The connection between the filtering of useable information and an engine's ability to run without unnecessarily wasting energy enables answering which observer does the best job in the sense of allowing for the smallest amount of energy dissipation \cite{CB}. Following this reasoning, one can {\em derive} from a purely {\em physical} argument, an information theoretic data compression strategy \cite{IB}, and hence find the algorithm that automates the ``intelligence" of the observer in information engines. (This algorithm trivially finds the optimal coarse graining for \LS's engine). 

This reasoning opens the door to using information engines not only to discuss the physical nature of information, but also to put adaptive information processing on a clear physical footing, and thereby provide a thermodynamic foundation to study the physics of intelligent observers. 
To that end, it is {crucial to know: what are the characteristics of observers that process information in this thermodynamically optimal way? }

{Here, we explore this question by analyzing a parameterized class of engines that arises naturally when \LS's model is generalized to partial observability (see sections \ref{partial} and \ref{tiltedBox}). This model is simple and generic, allowing us to isolate effects stemming from the mismatch between what the observer can perceive, and what the observer can do. The class of generalized partially observable \LS\ engines and their corresponding optimal observers, which we analyze here, serve as a canonical example to provide a basis for further considerations (see discussion at the end of section \ref{eta}). 
We find that parametric optimization of deterministic coarse grainings does not discover minimally dissipative observers. Optimal memories are intrinsically probabilistic. We compute them, and analyze their physical characteristics in sections \mbox{\ref{optimalObservers}--\ref{eta}}. The cost-benefit trade-off structure is analyzed in section \ref{optimalObservers}, and its connection to the computation of a rate--distortion function is explained. section \ref{Sec:opt_mem} provides a model for a physical implementation of probabilistic, minimally dissipative memories, and the thermodynamic efficiency these memories afford is examined in section \ref{eta}. }

\section{Information Engines}
Information engines alternate between two basic steps. First, information is acquired in a form that is stable over the remaining cycle. Then, this information is used to convert heat from an available heat bath to work output. Some models in the recent literature have focused only on the information-to-work step \cite{mandal2012work, strasberg2017quantum, stopnitzky2019physical}, neglecting the thermodynamics of the memory. In that way, one can arrange things to use information as fuel. We follow here instead the normal way of analyzing thermodynamic machines as cyclic processes. For information engines, this means that costs for running a meta-stable information carrier (memory) must be included in the engine's overall energy bill.

Therefore, the term {\em information engine} denotes a cyclic process, comprised of (a) an isothermal work-to-information transformation, and (b) an isothermal information-to-work transformation. The former represents a simple, mechanistic ``observer" inside the engine, acquiring and converting observable data to a physical representation thereof that is stable over the remaining cycle.  We refer to this as the {\em data representation step}. The latter uses a {\em work medium} to convert heat from the bath to work output via a given work extraction apparatus. We refer to this step as the {\em work extraction step}, because work is done by, i.e., extracted from, the engine.

Information is always context dependent, it is information {\em about} something. A classical bit of information corresponds to a meta-stable state of a system that can be in two, identifiably different, states. What that bit {\em means} depends on the context: if it is correlated with another system, or another degree of freedom in the same system, then it contains information about this other system, or this other degree of freedom. To use this information, there must be a read-out mechanism in place, some device that influences a sequence of events {\em depending} on the state of the physical information bit. Those two components are what give the abstract bit concrete meaning, and make it potentially useful.

In information engines, the work extraction device plays the role of the read-out mechanism. For example, in \LS's engine, the quasi-static movement of the divider can be arranged to lift a weight \cite{szilard}. It is the set of all used implements (in this case, divider, weight, and pulleys) that constitute the work extraction device. To function, the \LS\ engine requires knowledge of which side of the container is empty. A different work extraction device might require different information.  
We call all quantities that need to be known to extract work, {\em relevant}. It is clear from the discussion that the physics of the work extraction device define which quantity (or quantities) is (are) relevant.

The observer's memory must contain information about the relevant quantities. The ultimate information-to-work conversion limit of, on average, 1 bit to $kT \ln(2)$ joules can only be reached by the engine if the relevant variables are known with certainty. For that to be possible, the observable data has to contain all the information about the relevant variables, and the memory process has to retain that information. 

Engines for which the relevant variables are not fully observable need to infer those from the available data. Partial observability is a generic constraint faced by most real-world systems. It means that the observer inside the engine must solve an inference task. The efficiency of {\em partially} observable information engines depends upon how well the observer solves the inference task \cite{CB}. 

We may allow {\em generalized} information engines to run the two distinct processes of (a) forming the memory, and (b) extracting work, at two different temperatures, $T$ and $T'$, respectively \cite{CB}. This enables a combined treatment of heat engines and information engines. 

{\em Standard} information, upon which the literature has largely focused, are fully observable, and
are run isothermally. They have $T'=T$, and the relevant variables are either contained in the set of
observables, or fully reconstructable from it without uncertainty. Standard information engines are thus a
subset of the larger class of generalized, partially observable information engines, introduced in \cite{CB}.

The study of generalized, partially observable information engines enables the derivation of a constructive data compression algorithm from optimizing energy efficiency: maximizing the upper bound on the engine's net average work output, or, equivalently, minimizing the lower bound on dissipation, over all possible data representations yields a method for predictive inference \cite{CB}, known in lossy compression and machine learning as the {\em Information Bottleneck} \cite{IB} method. The Information Bottleneck approach has been generalized to dynamical and interactive learning \cite{Still-IAL09}, as well as to quantum information processing \cite{grimsmo2013quantum}. The generalized Information Bottleneck framework contains interesting classes of models as special cases \cite{Still-IBPI-2014, StillCruEl10}, some of which might explain some important aspects of neural processing and machine learning \cite{wiskott2003slow, creutzig2009past, tishby2015deep, shwartz2017opening}.

\LS's ideas paved the way for a physics based understanding of information. The core insight is that information is proportional to energy flows. It is thus logical that the efficiency of information {\em processing} can be measured in thermodynamic terms. The derivation of minimally dissipative observers, provided in 
\cite{CB}, opens the door to a physics-based understanding of the foundations of signal processing and learning, for which the physical characteristics of minimally dissipative observers need to be understood. We analyse this in sections \ref{optimalObservers}--\ref{eta}. Keeping in mind this bigger picture, we now review some aspects of the \LS\ engine, before introducing our canonical example in sections \ref{partial} and \ref{tiltedBox}. To build intuition, we then perform a naive parametric optimization of deterministic memories in section \ref{WE-mems}, before considering probabilistic memories.

\section{\LS\ Engine}
\label{Sbox}
In comparison to the Carnot process, \LS's engine does not require access to heat baths of different temperatures. In both cases, the isothermal expansion, in which work is produced by the engine, is identical, but to achieve volume reduction, two different strategies are used.  

In the Carnot process, the gas is compressed, a process which requires work. Isothermal, quasi-static compression of a one particle gas to half of the accessible volume requires that work be done on the gas in the amount of $kT \ln(2)$ joules. If the expansion of the gas were to take place at the same temperature, then this cost would annul any work done by the engine. The expansion must happen at a larger temperature, $T' > T$, for the engine to produce net work output. In this way, the engine converts a temperature difference to work.  

In the \LS\ engine, the isothermal compression is replaced by three steps: insert the divider in the center ($x=0$), measure the particle's $x$ position (from which the observer knows with certainty which side of the container is empty), and remember the empty side for the duration of the work extraction process. It is assumed that no work is necessary to insert the divider. Work is extracted at the same temperature---the \LS\ engine, in contrast to the Carnot process, does not appear to need a temperature gradient. 

However, the memory used by the \LS\ engine has to be implemented by some physical device, and the energy bill involved in running it must be taken into account. To memorize one of two outcomes, the memory must have at least two states, stable on the timescales needed. There are many possible implementations. 
The memory attains meaning by being coupled to degrees of freedom that allow for work extraction. \LS\ envisioned a mechanical switch, imagining that a weight could be attached to the moving divider via either of two pulleys, 
such that the weight gets lifted when the divider moves into the empty side of the container. One could say that the position of the mechanical switch constitutes the memory, but, in general, any bistable system could be used as memory.

\begin{figure}[ht]
\includegraphics[width=0.95\linewidth]{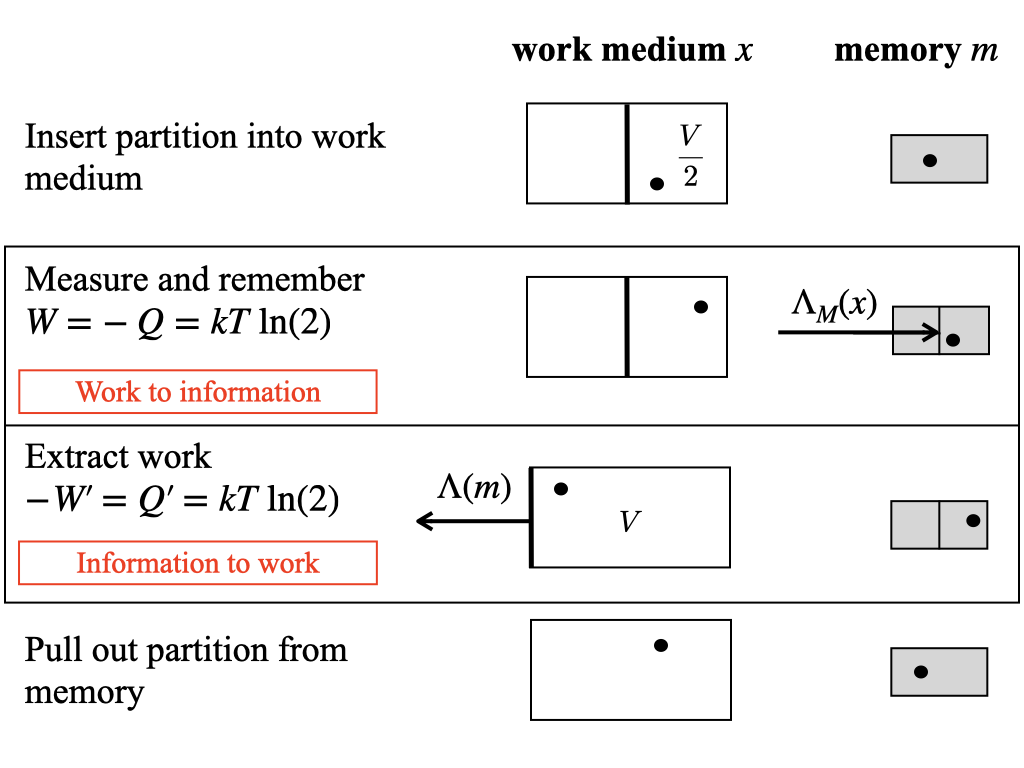}
\caption{ Sketch of \LS\ engine cycle, explicitly including another physical system to implement the memory---here another one particle gas (right column, gray). The protocol applied to the memory depends on the measurement of the particle's $x$ position, and is denoted by $\Lambda_M (x)$: if $x<0$, compress the memory such that $m<0$, and if $x>0$, such that $m>0$ (this case is sketched). Conversely, the work extraction protocol, $\Lambda (m)$, performed on the work medium (center), results in expansion of the gas towards the right, if $m<0$, or towards the left, if $m>0$ (sketched). Average work and heat are displayed for quasi-static transformations and $T'=T$.} 
\label{LSbox}
\end{figure}

A bistable memory can be modeled by a particle in a double well potential, as in \cite{Landauer1961}. The potential could be turned into a box potential with a barrier of infinite height and vanishing width. This means that the memory could be implemented by another one-particle gas in a container, together with a piston, as sketched in figure \ref{LSbox}. 

To ensure that the memory is in a specific state, the gas has to be compressed to half the volume---work is done on the gas, on average at least $kT \ln(2)$ joules. 

This simple argument alone shows that \LS's engine could never produce net work output. At best, \LS's engine can recover the energetic costs encountered by the observer for running the memory. Unless the memorizing is implemented at a {\em lower} temperature than the work extraction process.

We end up with an engine that, while producing no net work output when run at fixed temperature, nonetheless provides a concrete mechanism to {\em convert} between work and information. On average, the process that generates a data representation turns $kT \ln(2)$ joules into one bit of information, and the work extraction process turns one bit into $kT \ln(2)$ joules. In units of nats, which might be more natural for some, on average one nat can be converted into $kT$ joules and vice versa.

\subsection{Erasure protocols}
It has been discussed at great length in the literature that the costs associated with cyclically running a memory can be completely assigned to the erasing of that memory, by implementing a carefully designed protocol \cite{Landauer1961, bennett1982thermodynamics, sagawa2009minimal, parrondo2015thermodynamics, fahn1996maxwell, ouldridge2018power, barkeshli2005dissipationless}. This special protocol that contains a step in which the memory is reset to a certain state (i.e. erased) is more involved than our simple protocol. Here, we simply set the memory via isothermal compression, and later release the partition (figure \ref{LSbox}). In a protocol containing erasure, the setting of the memory is cleverly split apart into the erasing step and a step correlating the new state of the memory with the measurement outcome. 

The bottom line costs are identical for both protocols, namely at least $kT$ times the information captured in memory \cite{sagawa2009minimal, CB, parrondo2015thermodynamics, zurek1989algorithmic}. However, with the right choice of parameters, it can be made to look as if all the cost is encountered during the erasing step \cite{sagawa2009minimal}. Including erasure in the protocol is thus not a necessity, but rather an optional (ultimately arbitrary) choice. To go through the argument in detail, we generalize, in \mbox{appendix \ref{App-erasure}}, the erasure treatment from \cite{sagawa2009minimal}, which was introduced to elucidate precisely this point, and we compare it to the simpler set-and-release protocol considered here (figure \ref{LSbox}).

\subsection{Some variants of \LS's engine}
\label{LS-variants}
Many variants of \LS's engine have been explored in the literature, for example, containers with $K$ dividers and $N$ particles, e.g. \cite{song2019optimal} and references therein. Importantly, quantum \LS\ engines have been analyzed \cite{zurek1986maxwell,lloyd1997quantum,deffner2016quantum}, and it has been determined under what conditions the assumption is valid that sliding in the divider costs negligible energy \cite{zurek1986maxwell}.

It is instructive to calculate what happens if the divider is put off-center into the container. \LS\ considered this possibility. The volume then gets divided into a smaller volume, $\lambda V$, on one side, and a larger volume, $V - \lambda V$, on the other side, with $0 < \lambda \leqslant 1/2$. 

Thus, with probability $\lambda$, the particle will be found on one side, and with probability $1- \lambda$ on the other side of the divider. The work that can be extracted differs, depending on where the particle is. If it is in the smaller part, then more work can be extracted. 

In this paper, we adopt the convention that work done on the gas is positive, while work done by the gas is negative, and called {\em extracted} work. 
The work that can be extracted in the isothermal, quasi-static expansion of the one particle gas from initial volume, $V_i$ to final volume $V_f > V_i$ at temperature $T$, is proportional to the logarithm of the volume change, 
\begin{equation}
\label{W}
- W = kT \ln\left({V_f \over V_i} \right).
\end{equation}
In appendix \ref{thermo-basics} we explain this for readers unfamiliar with the matter. 

The total average extracted work, $-W_{\rm ex}$, is then composed of two terms: with probability $\lambda$, the gas expands from $\lambda V$ to $V$, contributing 
$kT \ln(1/\lambda)$ joules, and with probability $1- \lambda$, it expands from $V-\lambda V$ to $V$, contributing $kT \ln(1/(1-\lambda))$ joules: 
\begin{equation}
-W_{\rm ex}(\lambda) = kT \left[ \lambda \ln\left({1 \over \lambda}\right) + (1- \lambda) \ln\left({1\over 1-\lambda}\right) \right]. \label{w-lambda}
\end{equation}
We see immediately that this is the maximum amount of information the observer could have about which side of the divider is empty, because the probability of these events is given by $\lambda$ and $1-\lambda$, respectively. 

To state this more formally, define the random variable $U$ with two outcomes, $u$: if the $\{$larger, smaller$\}$ side of the container is empty, then let $u=\{0,1\}$. To write the probability of either outcome, $p(u) := p(U=u)$, we use the following notation: $p(u=0)$ denotes the probability that random variable $U$ has the outcome $u=0$. Here, the probability that the larger side of the container is empty is $p(u=0)=\lambda$, while $p(u=1)=1-\lambda$, and the entropy is $H[U] = - \lambda \ln\left({\lambda}\right) - (1- \lambda) \ln\left({1-\lambda}\right)$ \footnote{Notation: entropy and information are functionals of probability distributions. But the information theory literature \cite{Cover1991} often adopts the following shorthand notation, which we use: $H[U] \equiv -\left\langle \log\left[ {p(u)}\right] \right\rangle_{p(u)}$ for entropy, $H[U|M] \equiv  -\left\langle \log\left[ {p(u|m)}\right] \right\rangle_{p(m,u)}$, for conditional entropy, and for mutual information $I[M,U] \equiv \left\langle \log\left[ {p(m,u) \over p(m)p(u)} \right] \right\rangle_{p(m,u)} \label{MI}$. Brackets, $\langle \cdot \rangle_p$ denote averages over the probability distribution $p$.}. Thus, the average extractable work is $-W_{\rm ex}(\lambda) = kT H[U]$.

By definition of mutual information, any other random variable (for example a memory, $M$) could carry information up to $I[M,U] = H[U] - H[U|M]  \leqslant H[U]$ (because classical conditional entropy is a non-negative quantity, i.e. $H[U|M] \geqslant 0$).

To maximize average work done over the parameter $\lambda$, one finds $\lambda = 1/2$. Placing the divider anywhere but in the middle does not increase the engine's average work output. 

A similar calculation leads to the insight that if the measurement is corrupted by error, or the memory is corrupted by a noise process, then one should leave a residual space for the particle, the relative volume fraction of which is given by exactly the error probability \cite{sagawa2012thermodynamics}. 

To see this, for simplicity assume the divider is in the middle of the container, at position $x=0$, and assume the container has volume V, with unit transverse area, so that it extends in $x$-direction between $x=-V/2$ and $x=V/2$. A two-state memory will be formed and then used to adjust the work extraction contraption as follows: if $m=-1$, then the divider will move quasistatically to the right until it reaches position $x=V(1/2 -\gamma)$, whereby a residual fractional volume $\gamma V$ remains on the right side of the divider, with $1/2 \geqslant \gamma > 0$. Symmetrically, for $m=1$, the divider moves to position $x=V(\gamma-1/2)$.

The memory's state $m=-1$ is assigned to the event that the particle is to the left of the divider, i.e. $x<0$, but with probability $q$ an error is made, and the memory state ends up in state $m=1$, hence $p(m=1|x < 0) = q$. Symmetrically, for the other side: $p(m=-1|x > 0) = q$. 

Whenever an error is made, the action taken in response will lead to the gas being compressed, rather than expanded, and this occurs with probability $q$. To avoid compression to zero volume, the residual volume $\gamma V$ is left. Work will be done on the gas with probability $q$, in the amount of $kT \ln(1/2 \gamma )$. Whenever no error is made, work is extracted in the amount of, $kT [\ln(2) + \ln(1-\gamma)]$ joules. The extractable work is reduced because of the residual volume ($\ln(1-\gamma)$ is negative). Altogether, we can write down the total average extracted work, 
\begin{equation}
\!-W_{\rm ex}(\gamma) = kT \left[ \ln(2) + q \ln(\gamma) + (1-q) \ln(1-\gamma) \right], \label{Wextract}
\end{equation}
and maximize over $\gamma$, to find $\gamma = q$.

The maximal extractable work is thus proportional to the information the memory captures about which side of the container is empty. To see that,  
let the random variable $U$ have the two realizations $\{ -1, 1\}$, where $u=-1$ denotes that the right side is empty, while $u=1$ denotes that the left side is empty. Then $p(u) = 1/2$, $H[U]=\ln(2)$, 
\begin{eqnarray}
p(u|m) &=& 
\begin{cases}
1-q, & u=m \\
q, & {\rm else}
\end{cases} 
\end{eqnarray}
and $I[M,U] = \ln(2) + q\ln(q) + (1-q) \ln(1-q)$.
Hence, 
\begin{equation}
\!-W_{\rm ex} = kT I[M,U].
\end{equation}

\subsection{Partially observable generalized \LS\ engines}
\label{partial}
Measurement error is not the only mechanism that can limit available information. Real world observers typically face constraints that limit how much information is available and can be used. Therefore, there are many situations in which lowering measurement noise is insufficient. The case in which an observer can directly measure all quantities that need to be known to extract work is a special case in the larger class of situations. The more general class of observers has access only to observables that are correlated with relevant quantities, but not necessarily in a one-to-one relationship. In this general case, the observer needs to infer the relevant quantities from the observables. Not all of the information the observer keeps in memory about the observables is necessarily informative about these relevant quantities. The total memory can thus be split into relevant, useful information, plus irrelevant, useless information. The amount of relevant information stored in memory depends on the data representation strategy used to map observable data to memory states, which is characterized by the conditional distributions $p(m|x)$. The amount of heat dissipated in a cycle, and the thermodynamic efficiency of the resulting engine depend on this data representation \cite{CB}. Data representations that allow for the smallest possible dissipation are those that capture as much relevant information as possible while retaining as little irrelevant information as possible, i.e., those that are best at predicting the quantities of interest from the available observations \cite{CB}. 

One way to picture a scenario involving partial observability is to ask what would happen if \LS's protocol was changed to measuring and making the memory {\em before} inserting the divider. The observer could then use the memory to decide where to insert the partition, to maximize expected work output. The observer would have to {\em predict} the particle's location at the time of insertion, taking the delay between measurement and insertion into account. For that, the observer must estimate the particle's velocity. However, the velocity  is not directly observable, only the particle's $x$ position is. The observer has to infer velocity from differences in position. The involvedness of this model would obfuscate the crucial points illuminating the trade-off between thermodynamic cost of a memory and thermodynamic gain derivable from the memory. In particular, the delay dependent estimation error is convoluted with the error stemming from the {\em mismatch} between what the observer can perceive and what the observer can actively change in the observed system. 
It is important to realize that while partial observability can manifest in some cases through measurement error alone, in the more general case it is due to this mismatch, representing a fundamental limitation. In this general case, the partial observability constraint cannot be eliminated, not even by decreasing the measurement error to zero.

\begin{figure}[h]
\centering
\includegraphics[width=0.95\linewidth]{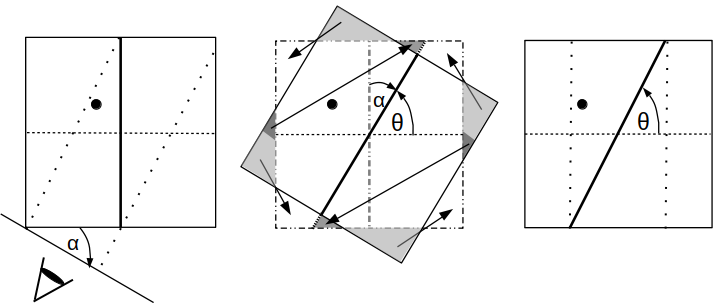}
\caption{Equivalence (center panel) between (1) tilting the observer's viewing angle $\alpha$ (left panel), and (2) inserting the divider at an angle $\theta = \pi/2 - \alpha$, while the particle's $x$ position is observable (right panel). Both alternatives result in equivalent partially observable \LS\ engines. }
\label{fig:TiltedView}
\end{figure}

To isolate and study the consequences of this mismatch, a much simpler, perhaps the simplest, way to visualize an example is sketched in figure \ref{fig:TiltedView}. 
Imagine that the observer part of the \LS\ engine measures the particle's position projected onto an axis tilted away from the $x$-axis by a viewing angle, $\alpha$ (left drawing). 

For the observer, this results in a situation with three distinct areas. 
There are two areas where knowing the particle location determines with probability one which side of the container is empty. Observing the particle in the central area (between the dotted lines sketched in the left and right panels of figure \ref{fig:TiltedView}), however, leaves some uncertainty.

An equivalent example results from inserting the divider in the original \LS\ box at an angle, $\theta$, while leaving the observer's viewing angle unchanged. Figure \ref{fig:TiltedView} shows the transformation between the two alternatives (middle drawing). The geometry with the tilted divider (right drawing) is slightly more intuitive, because the probability of finding the left side empty, given the particle's precise $x$ position, follows the shape of the divider,
see equation (\ref{pleft}) below.

\section{Tilted divider \LS\ box}
\label{tiltedBox}
\begin{figure}[h]
\centering
\includegraphics[width=0.85\linewidth]{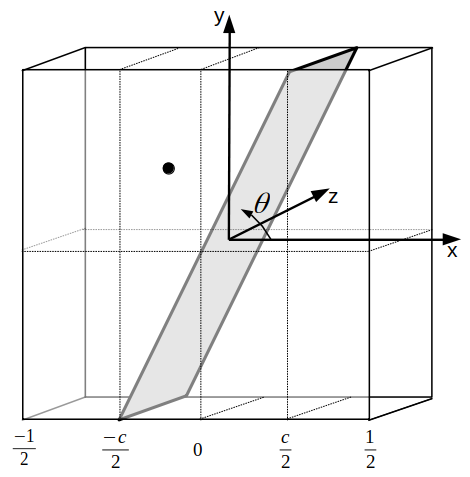}
\caption{Sketch of work medium in our model: one particle gas with a divider inserted at angle $\theta$, introducing an uncertain region, $c=\cot(\theta)$.}
\label{pos}
\end{figure}

Inserting the divider into the container at an angle is a minimal change to \LS's engine that yields a partially observable information engine, given that only the $x$ position of the particle is observable. 
The sketch in figure \ref{pos} shows such a modified \LS\ box, drawn in three dimensions, centered on the origin. The angle, $\theta$, between the $x$-axis and the divider parameterizes a family of partially observable information engines.
\LS's original engine is recovered for $\theta = \pi/2$. 

For simplicity, assume fixed unit length of the container in both $x$- and $y$-direction. The divider then touches the walls at $x=\pm \cot(\theta)/2$ (see figure \ref{pos}). The uncertain region is the area in between. In this area, the particle's $x$ position is correlated with which side of the divider is empty, but not in perfect one-to-one correspondence. 
The size of the uncertain region is $c\equiv \cot(\theta)$, vanishing for $\theta = \pi/2$. The uncertainty in this region is not attributable to measurement noise, but rather stems from the structural limitation. Even perfect, noise-free, measurements would not solve the issue. 

For simplicity of the exposition, we limit the angle between $\pi/4 \leqslant \theta \leqslant \pi/2$ in the main text, and include the less interesting range $0 \leqslant \theta \leqslant \pi/4$ in appendix \ref{flat-tilt-boxes}. In appendix \ref{comp-varyL-boxes}, we also discuss that an alternative, equivalent parametrization would be to fix the angle, and vary the length of the container in $x$-direction. The relative length of the certain regions compared to the uncertain region determines the amount of relevant, useable information contained in the measurement.
 
When the particle's $x$ position is recorded to the right of the central region, i.e., $x \geqslant c/2$, then we know for sure that the left part of the box is empty. Conversely, if $x \leqslant -c/2$, then the right half of the box must be empty. As before, let the outcomes of the random variable $U$ be $u=\{-1,1\}$, whenever the $\{$right, left$\}$ part of the box is empty. 

Which side of the container is empty is precisely what the observer needs to know to extract work. The random variable $U$ is thus the relevant variable here. For all \mbox{$x \in [c/2, 1/2]$} we know with certainty that the left side of the box is empty, i.e. $p(u=1|x)=1$, and similarly for $x \in [-1/2, -c/2]$ the right half must be empty, $p(u=1|x)=0$. However if the particle is observed in between, $-c/2<x<c/2$, we can only make a probabilistic guess, and the probability of either side being empty depends on $x$ linearly: $p(u=\pm 1|x) =  {1\over 2} \pm {x\over c}$. In summary, we have:
\begin{eqnarray}
\label{pleft}
p(u=1 |x) &=& 
\begin{cases}
0, & x \leqslant -c/2 \\
{1\over 2} + {x\over c}, & - c/2 < x < c/2 \\
1, & x \geqslant c/2
\end{cases}
\end{eqnarray}
with $c = \cot(\theta)$.

We always have a choice regarding how to represent available data. For the original \LS\ engine, it is clear how to do that. Without a tilt, it is self-evident to coarse grain the particle's $x$ position into the regions left and right of the divider, i.e. choose a
map from $x$ to $m$ such that $m=-1$, $\forall x<0$, and $m=1$, $\forall x>0$, that is set $m={\rm sign}(x)$. 

With a tilted divider, it is less obvious how to choose the optimal data representation. The finer we grain the particle's $x$ location, the more useful information we can capture. But we also require more energy to run a more elaborate memory. If the memory is prepared at a lower temperature, $T$, than the temperature at which the information is turned into work, $T'$, then there must be a natural trade-off between the cost incurred for a more detailed memory vs the added gain. The optimal data representation must therefore depend on the temperature ratio, \mbox{$\tau = T'/T \geqslant 1$}. 

Access to two heat baths at different temperatures can be introduced to the \LS\ engine by isentropic compression and expansion of the container along the $z$-axis. The transformations applied to change the temperature need to be such that correlations between memory and work medium are not changed in the process. The depth of the container in $z$-direction can be used to change the volume for the purpose of changing the temperature of the gas, while keeping the divider geometry fixed. 

\subsection{Relevant information in memory}
As before, imagine that another physical system serves as memory. Let that memory have $K$ distinct states $m=1,\dots, K$. The information that the memory captures about the observable,
\begin{equation}
I_{\rm mem} \equiv I[M,X] = \left\langle \ln\left[ {p(m|x) \over p(m)} \right] \right\rangle_{p(m,x)}
\end{equation}
depends on the conditional probability distributions $p(m|x)$ that characterize the stochastic map from observables to memory states. If this map contains no randomness, then the conditional entropy is zero, $H[M|X]=0$, and $I[M,X] = H[M]$. We call these assignments {\em deterministic}. 

How much of the memorized information is predictive of the relevant quantity? 
The relevant information that can be used to extract work is the mutual information retained in memory about which side of the box is empty:
\begin{equation}
\label{Irel-def}
I_{\rm rel} \equiv I[M,U] = \left\langle \ln\left[ {p(u|m) \over p(u)} \right] \right\rangle_{p(m,u)},
\end{equation}
where $u$ is inferred from $m$ by calculating 
\begin{equation}
p(u|m) = \langle p(u|x) \rangle_{p(x|m)}, \label{pum_1} 
\end{equation}
with probability density $p(x)=1/L$, conditional density $p(x|m) = p(m|x) p(x) / p(m)$ and \mbox{$p(m) =  \langle p(m|x) \rangle_{p(x)}$}. (Since we chose the length of the container in $x$-direction to be $L=1$, $p(x)$ will not show up in the following calculations. In appendix \ref{comp-varyL-boxes} we discuss the $L\neq 1$ case). In equation (\ref{pum_1}) we made use of the fact, that when the raw data is {\em given}, then the probability of $u$ does not depend on the memory state, i.e. $p(u|m,x) = p(u|x)$. Recall that the data representation chosen by the observer is characterized by $p(m|x)$, i.e. by the map from observable data to memory. It is important to remember that relevant information is a functional of $p(m|x)$. It also is a functional of $p(u,x)$, which, in turn, depends on the geometry of the box, that is, on the tilt angle.

The remainder of the memorized information cannot be used and is thus irrelevant:
\begin{equation}
I_{\rm irrel} \equiv I_{\rm mem} - I_{\rm rel}.
\end{equation}
Irrelevant information is non-negative, because $p(u|m,x) = p(u|x)$ implies that 
\begin{eqnarray}
\!\!\!\!\!\!\!\!\! I_{\rm irrel}
&=& \left\langle\ln{\left(p(m,x|u) \over p(m|u) p(x|u)\right)}\right\rangle_{p(m\!,x\!,u)} \!\!\! \equiv I[M,X|U].
\end{eqnarray}
Since conditional information, $I[M,X|U]$, is non-negative, we have $I_{\rm irrel} \geqslant 0$.

For the same reason, relevant information captured in memory can never exceed the total information, $I[U,X]$, that the particle's $x$ position contains about which side of the container is empty: $I[U,M] \leqslant I[U,X]$. 

To calculate $I[U,X] = H[U] - H[U|X]$ as a function of the tilt angle, $\theta$ (and thus as a function of the size of the uncertain region $c$),
first note that due to symmetry, $p(u) = 1/2$, and thus $H[U] = \ln(2)$.

Because of the natural logarithm, information is measured in nats. Thoughout the paper, depending on context, we sometimes divide by $\ln(2)$, to have information in units of shannon [Sh]. Colloquially, the name for the unit shannon is often replaced by bit, language we use throughout the paper.

Next, to calculate $H[U|X]$, we need the conditional probabilities, 

\begin{eqnarray}
p\left(u \Bigg| x\in \left[-{1\over 2}, -{c \over 2}\right]\right) &=& 
\begin{cases}
1, & u=-1 \\
0, &u=1
\end{cases} \\
p\left(u \Bigg| x\in \left[-{c\over 2}, \;\;\;{c \over 2}\right]\right) &=& 
\begin{cases}
{1\over 2} - {x\over c}, & u=-1 \\
{1\over 2} + {x\over c}, &u=1
\end{cases} \\
p\left(u \Bigg| x\in \left[\;\;\,{c\over 2}, \;\;\;{1 \over 2}\right]\right) &=& 
\begin{cases}
0, & u=-1 \\
1, &u=1
\end{cases}.
\end{eqnarray}
Inserting these into $H[U|X] = -\left\langle \ln\left[ {p(u|x)}\right] \right\rangle_{p(u,x)}$ gives 
\begin{equation}
H[U|X] = {c \over 2}. 
\label{HUgX}
\end{equation}
(The detailed calculation is provided in appendix \ref{AppA}.) 
Therefore, the particle's $x$ position contains information about which side of the box is empty in the amount of
\begin{equation} \label{Imax}
I[U,X] = \ln(2) - {c \over 2}  \equiv I^{\rm max}_{\rm rel}(c). 
\end{equation}
This is the maximum amount of relevant information that the observer could possibly extract from the measurement, capture in memory, and turn into work:
\begin{equation}
\label{upper-bound}
I_{\rm rel} \leqslant I^{\rm max}_{\rm rel}(c).
\end{equation}

\subsection{Work extraction from relevant information}
\label{workextraction}
Let us run the following memory-dependent work extraction protocol: 
\begin{enumerate}[label=(\alph*)]
\item Isolate the container from the heat bath and reduce the container's dimension in $z$-direction, thereby changing the volume of the entire container from $V$ to $V'\leqslant V$, increasing the temperature by the factor $\tau = T'/T$. \label{s1}
\item Connect to a heat bath at temperature $T'$. Extract work in an isothermal transformation in which the wall moves into the direction of the space that is estimated to be empty with highest probability, given the state of the memory, while leaving an optimal residual volume. \label{s2}
\item Pull the divider out. \label{s3}
\item Isolate the container from the heat bath and increase the container's dimension in $z$-direction to change the volume of the entire container from $V'$ to $V$ in order to lower the temperature back to $T$. \label{s4}
\item To close the cycle and prepare for the next cycle, re-insert the divider.  \label{s5}
\end{enumerate}
How much work does this protocol extract, on average? 

First we analyze the two isentropic processes, steps \ref{s1} and \ref{s4}. The isentropic compression in step \ref{s1} reduces the volume of the entire container from $V$ to $V'$.
Because the process does not change the geometry, the particle continues to reside in one half of that volume. Therefore, the volume the gas occupies changes from an initial volume of $V_i=V/2$ to a final volume of $V_f=V'/2$. This causes the temperature to change from the initial temperature $T_i = T$ to $T_f= T'$. The adiabatic condition relates volume change to temperature change, 
\begin{eqnarray}
{V_i \over V_f } = \left( {T_f \over T_i} \right)^{3 \over 2}, \label{adiabatic-eq}
\end{eqnarray}
(it is explained in appendix \ref{TB-2}). Evaluating it gives: 
\begin{equation}
V / V' = ( T' / T )^{3 \over 2}. \label{isentropic-compression-ratios}
\end{equation}

During the isentropic expansion in step \ref{s4}, the gas occupies the entire volume of the container, which changes from $V'$ to $V$. The starting temperature is $T_i= T'$. We evaluate equation (\ref{adiabatic-eq}) for this transformation, to get ${V' / V} = \left({T_f / T'} \right)^{3 \over 2}$. Together with equation (\ref{isentropic-compression-ratios}), this tells us the final temperature: $\left({T_f / T'} \right)^{3 \over 2} = ( T / T' )^{3 \over 2} \Rightarrow T_f = T$.
We obtain the same temperature which we had at the beginning of the work protocol. 

The work done on the gas during isentropic compression is compensated by the work during isentropic expansion (this is explained in appendix \ref{W-cancels}). 

The average work that can be extracted from the one particle gas in our engine during the isothermal expansion at temperature $T'$ in step \ref{s2}, depends on the accuracy with which $u$ can be inferred from $m$, via $p(u|m)$, equation (\ref{pum_1}). 
This inference may result in an error $q(m)$, defined as follows. Let $\hat{u}$ be the true value of $u$. Then the probability 
$p(u=\hat{u}|m) = 1-q(m)$. This means that with probability $1-q(m)$ the gas expands and does work against the piston, but with probability $q(m)$ the piston does work on the gas, which is compressed. Therefore, we must leave a residual volume. The optimal value for this residual volume can be computed similarly to the calculation around equation (\ref{Wextract}), but now the average amount of work that is extracted, given a memory state $m$, depends on $q(m)$, and thus the residual volume $\gamma(m)$ must be chosen accordingly. With probability $1-q(m)$ the gas is expanded from $V'/2$ to  $V'-\gamma(m)V'$, and with probability $q(m)$ it is compressed from $V'/2$ to  $\gamma(m)V'$. Thus,
\begin{eqnarray}
\!\!\!\!\!\! - W'(m) &=& kT'\left[ [1-q(m)] \ln\left({[1-\gamma(m)]V' \over {V'/2}}\right) \right.\\
&& \;\;\;\;\;\;\;\; \left.- q(m) \ln\left({V'/2 \over \gamma(m)V'}\right) \right] \notag\\
&=& kT'\bigl( \ln(2) + [1-q(m)] \ln\left[1-\gamma(m) \right] \\
&& \;\;\;\;\;\;\;\;\;\;\;\;\;\;\;\; + q(m) \ln\left[{\gamma(m)}\right]\bigr). \notag
\end{eqnarray}
As before, this is maximized by choosing $\gamma(m)=q(m)$, which gives an average work, averaged over all memory states, of
\begin{eqnarray}
- W' &=& kT'\left( \ln(2) + \sum_m p(m) \sum_u p(u|m) \ln\left[p(u|m) \right]  \right) \notag  \\
&=&  kT' (H[U] - H[U|M]) =  kT' I_{\rm rel}. \label{WeqIrel}
\label{Wout}
\end{eqnarray}
We have thus verified that the work extraction protocol extracts an average amount of work proportional to the relevant information kept in memory \cite{CB}. 

The isothermal transformations in this work protocol are performed quasi-statically, in such a way that the system remains in equilibrium throughout. The work calculated in equation (\ref{WeqIrel}) is thus an upper bound on how much work can be extracted from the tilted divider \LS\ box, on average.

\section{Deterministic memories, parametrically optimized for work extraction}
\label{WE-mems}
The absolute maximum extractable work depends on the geometry of the box, that is, on $c$, see equations (\ref{Imax}) and (\ref{upper-bound}). 
But how much relevant information is retained in memory depends on the choice of data representation, via $p(m|x)$. Different representations capture different amounts of relevant information. 

To build intuition, we now explore what fraction of this maximally retainable relevant information can be extracted with deterministic $K$-state memories (coarse graining the $x$-axis into $K$ regions), where $K$ ranges from 2 to 5. 

\begin{figure}[ht]
\centering
\begin{subfigure}{.43\linewidth}
    \centering
    \includegraphics[height=1.8in]{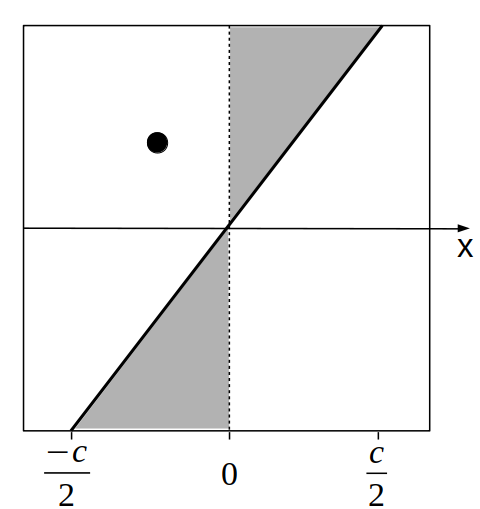}
    \caption{Deterministic 2-state memory parametrization.}
    \label{fig:2state_err}
\end{subfigure}
    \hspace{1em}
\begin{subfigure}{.43\linewidth}
    \centering
    \includegraphics[height=1.8in]{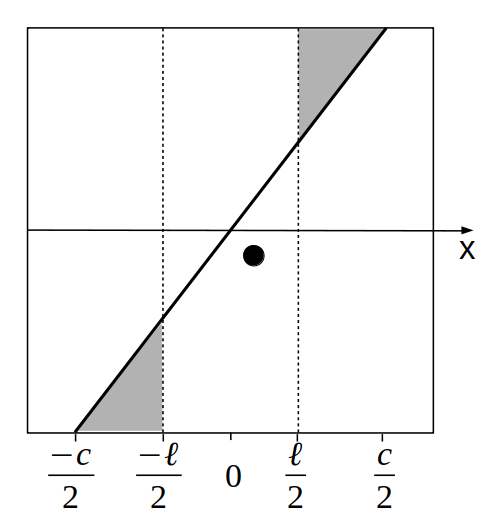}
    \caption{Deterministic 3-state memory parametrization.}
    \label{fig:3state_err}
\end{subfigure}
\bigskip
\begin{subfigure}{.43\linewidth}
  \centering
  \includegraphics[height=1.8in]{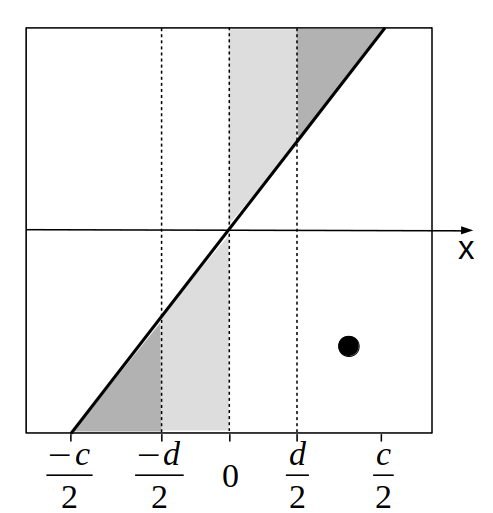}
  \caption{Deterministic 4-state memory parametrization.}
  \label{fig:4state_err}
\end{subfigure} 
\hspace{1em}
\begin{subfigure}{.43\linewidth}
    \centering
    \includegraphics[height=1.8in]{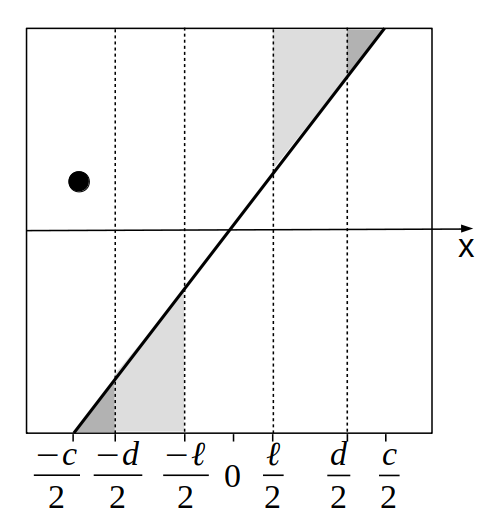}
    \caption{Deterministic 5-state memory parametrization.}
    \label{fig:5state_err}
\end{subfigure}
\caption{Parameterized deterministic partitions of the observable into memories with $K = 2, \dots, 5$ states. Shaded regions are proportional to the error, $q_K(m)$.}
\label{grrr}
\end{figure}
Intuitively, one expects that coarse graining should respect the symmetry of the problem.
For two states, the best way to coarse grain is to assign $m = {\rm sign}(x)$, as in the original \LS\ engine. For three states, the choice is more complicated, but can be parameterized by one parameter, $\ell$, that measures the size of the center partition. Four states can also be parameterized by just one parameter, due to symmetry. Five memory states are parameterized by two parameters. These parametrizations are shown in figure \ref{grrr}. 

\subsection{Two-state memory} 
\label{2s}
The deterministic two-state memory is given by
\begin{eqnarray}
p\left(m|x\right) &=& 
\begin{cases}
1, & {\rm if} \;\; m={\rm sign}(x) \\
0, & {\rm otherwise}
\end{cases}. 
\end{eqnarray}
The error probability has to be the same for each memory state, due to symmetry, i.e. $q_2(m)\equiv q_2$. The probability of error can be read off from the geometry of the box. It is $q_2 = c / 4$, which ranges from $q_2=0$ at $\theta=\pi/2$ to $q_2=1/4$ at $\theta=\pi/4$. Therefore, the optimal strategy for the isothermal expansion at $T'$ in step \ref{s2} is to leave a residual volume fraction of $\gamma=q_2 = c/4$. We then extract, on average
\begin{eqnarray}
-{W'}^{(2)}&=& k T' \left[  \ln(2) - h(c / 4) \right], \label{W2}
\end{eqnarray}
where we use a non-negative entropy function with 
\begin{eqnarray}
h(x) &\equiv& - \left(1- x \right) \ln\left(1-x\right) - x \ln(x) \geqslant 0, \label{Entq}
\end{eqnarray}
defined for $x \in (0,1)$, and \mbox{$h(x=0)=h(x=1)=0$}. 

\begin{figure}[t]
\centering 
\includegraphics[width=0.8\linewidth]{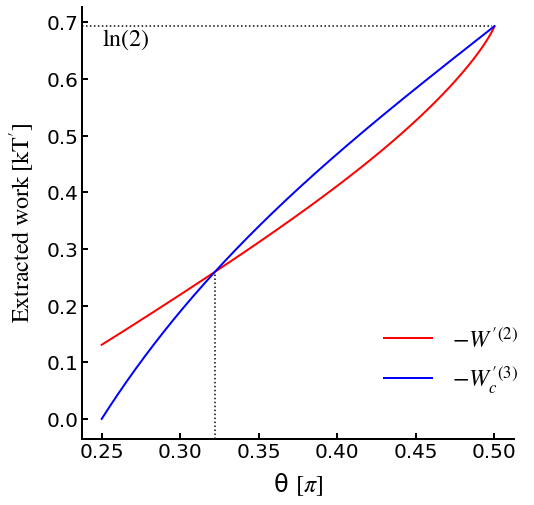}
\caption{ Average extracted work as a function of $\theta$. Red line ($-{W'}^{(2)}$, equation (\ref{W2})): using the deterministic two-state memory. 
Blue line ($-{W'}^{(3)}_c$, equation (\ref{W3})): using the three-state memory with $\ell=c$.}
\label{workcomp}
\end{figure}

The average extracted work is positive, and increases as $\theta$ increases from $\pi/4$ to $\pi/2$. The less tilted the divider, the closer to $kT'\ln(2)$ can be extracted (see figure \ref{workcomp}).

\begin{figure}[h]
\centering 
\includegraphics[width=0.8\linewidth]{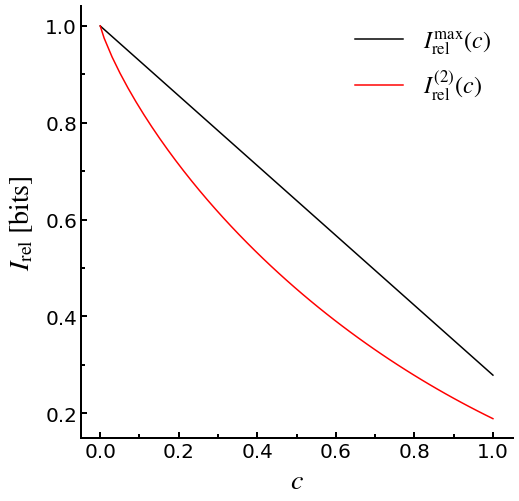}
\caption{ Relevant information retained by the deterministic two-state memory, $I_{\rm rel}^{(2)}(c)$, equation (\ref{Irel2}), (red), as a function of $c$. Compared to maximally extractable relevant information, $I_{\rm rel}^{\rm max}(c)$, equation (\ref{Imax}), (black).}
\label{2state-Irel}
\end{figure}

Let us be incredibly pedantic and double-check that the relevant information is indeed $\ln(2) - h(c / 4)$. Due to symmetry, we have 
\mbox{$p(m)=1/2$}, and 
\mbox{$p(u) =1/2$}, hence $H[U] = \ln(2)$. The conditional probabilities are 
\mbox{$p(u\neq m|m) = c/4$}, and \mbox{$p(u=m|m) = 1-c/4$}. This results in a conditional entropy of
\begin{eqnarray}
H[U|M] &=&-\!\!\!\!\!\! \sum_{m \in \{-1,1\}} {1\over 2} \sum_{u \in \{-1,1\}} \!\!\!\!\!\! p(u|m) \ln[p(u|m)]\\
&=& - \left(1-{c \over 4}\right) \ln\left(1-{c \over 4}\right) - {c \over 4} \ln\left({c \over 4}\right) \notag \\
&=& h(c / 4),
\end{eqnarray}
and hence for the relevant information
\begin{eqnarray}
I_{\rm rel}^{(2)}(c) &=& I[M,U] = H[U] - H[U|M] \\
&=&  \ln(2) - h(c / 4). \label{Irel2}
\end{eqnarray}
While the total relevant information available in the particle's $x$ position, equation (\ref{Imax}), decreases linearly with the size of the uncertain region, $c$, the amount of relevant information captured by this coarse graining decreases as a convex function, see figure \ref{2state-Irel}. The difference, which is a measure for lost information, will thus have a maximum, but rather than the absolute loss, we are interested in the relative loss, namely how much information is lost as a fraction of how much there is to be summarized.

Equivalently, we can plot the retained relevant information as a fraction of the maximum relevant information, i.e., $I_{\rm rel}^{(2)}(c)/I_{\rm rel}^{\rm max}(c) = [\ln(2) - h(c / 4)]/ (\ln(2) - {c / 2})$. This is done in figure \ref{optimalIfrac} (blue triangles), where it is compared to memories with more states, which we will address in the following paragraphs. 

\subsection{Three-state memory} \label{3s}
Compare this to a three state memory. To begin, let us choose the simple three-state coarse graining that maps each certain region to one state ($m = 1$ if $x \geqslant c/2$ and $m = -1$ if $x \leqslant - c/2$), and maps the uncertain region to a third state ($m=0$ if $-c/2 < x < c/2$). 
The error for states $m=\pm 1$ is zero, $q_3(m=\pm 1)=0$. Every time the memory is in one of these states, we can thus let the volume expand by a factor 2. The total probability that one of these two cases occurs is $1- c$. But for $m=0$, the error is $q_3(m=0)=1/2$, and thus we cannot do anything to extract work, on average.

Note that the probability of not being able to do anything, $c$, grows to $1$, as $\theta$ gets closer to $\pi/4$, which makes the memory effectively a one state memory, i.e., for $\theta=\pi/4$, when there are no certain regions, no information is contained in this coarse graining. 

The average work the observer can extract using this memory is
\begin{equation}
-{W'}^{(3)}_c = k T' \ln(2)(1 - c), \label{W3}
\end{equation}
which is zero for $\theta=\pi/4$, and increases to $kT' \ln(2)$ as $\theta$ approaches $\pi/2$.  

Because of the symmetry, memory state $m=0$ cannot be used to extract work, on average. Hence, this memory state is akin to a ``waste basket" of width $c$: when the $x$ position appears in this region, we basically throw the measurement away. This helps to decrease the error made when the memory is in the other states to zero. But it is not necessarily the way to get the most work out. Just how bad this strategy is depends on the geometry of the box, i.e. the width of the waste basket. There is a value, $\theta^*$, such that more work can be extracted with the two state memory for all $\pi /4 \leqslant \theta<\theta^*$: \mbox{$-{W'}^{(2)}>-{W'}^{(3)}_c$}. This value is determined by setting equation (\ref{W2}) equal to equation (\ref{W3}), and comes out to roughly $0.32 \pi$ (see figure \ref{workcomp}). 

Let us see if the observer can do better with three memory states. Let $\ell$ be the width of the interval assigned to the waste basket memory state, $m=0$. In general, $\ell$ can differ from $c$. Choosing $\ell > c$ decreases the amount of relevant information captured. If we assign a smaller interval to the $m=0$ state by choosing $\ell < c$, then the probability of not be able to do anything is reduced. However, when we have a signal, there is now a residual error. This error can again be read off from the geometry of the box. 

It is given by the area of the triangle on the other side of the divider, divided by the area that corresponds to state $\pm 1$, respectively (see figure \ref{fig:3state_err}). The area corresponding to states $m=\pm 1$ is $(1-\ell)/2$, respectively. The area of the triangle is $(c-\ell)^{2} / 8c$. Due to symmetry the error for both states is the same, $q_3(m=-1)= q_3(m=1) \equiv \epsilon (\ell, c)$, with
\begin{eqnarray}
\label{eps}
\epsilon (\ell, c)&=& {(c-\ell)^{2} \over 4c (1-\ell)}.
\end{eqnarray}
The error for state $m=0$ remains $q_3(m=0)=1/2$. The relevant information captured by this memory is $I[U,M] = H[U] - H[U|M] = \ln(2) - H[U|M]$, with 
\begin{eqnarray}
H[U|M] &=& \ell \ln(2) - (1-\ell)  \Bigl[\epsilon(\ell, c) \ln[ \epsilon(\ell, c)]  \\
&& +[1-\epsilon(\ell, c)] \ln([1-\epsilon(\ell, c)]) \Bigr] \notag \\
&=& \ell \ln(2) + (1-\ell)\,h(\epsilon(\ell, c)),
\end{eqnarray}
and thus
\begin{eqnarray}
I_{\rm rel}^{(3)}(\ell) &=& (1-\ell) \big[\ln(2) - h(\epsilon (\ell, c))\big]. \label{Irel3}
\end{eqnarray}
If the residual volumes in the work extraction protocol are properly adjusted to match the error probability, then the average extractable work is $-W^{(3)}= kT' I_{\rm rel}^{(3)}(\ell)$. 

\begin{figure}[ht]
\centering 
\begin{subfigure}{\linewidth}
\centering
\includegraphics[width=.53\linewidth]{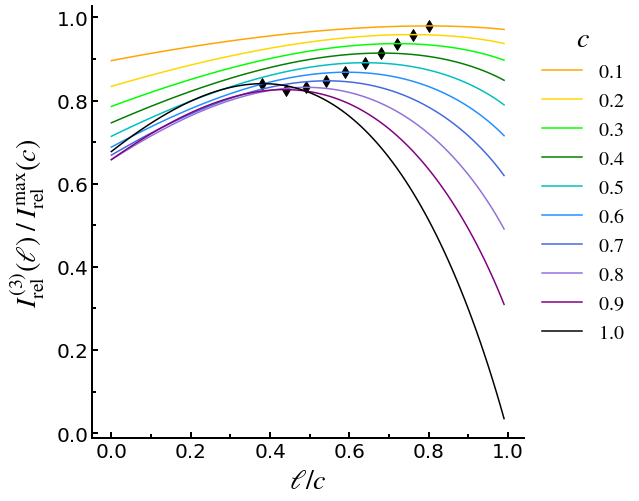}
\includegraphics[width=.43\linewidth]{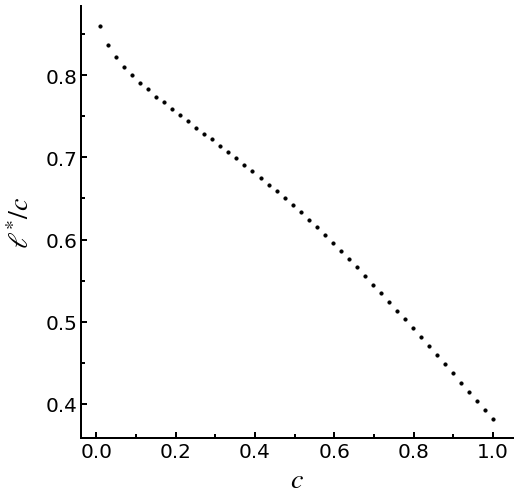} 
\end{subfigure}
\caption{Normalized average extracted work for the three-state memory as a function of $\ell/ c$ (left). Relative waste basket length $\ell^*/c$ for the three-state memory that maximizes over $\ell$ the relevant information retained, equation (\ref{Irel3}), and thus the potential thermodynamic gain (right). Plotted as a function of $c$.}
\label{3-W-l}
\end{figure}

We plot $I_{\rm rel}^{(3)}(\ell)$, normalized by $I_{\rm rel}^{\rm max}(c)$, as a function of $\ell/c$ in figure \ref{3-W-l}, for ten different values of $c$.

There is an optimal waste-basket length $\ell^*(c)$ associated with the maximum extractable work (diamonds in figure \ref{3-W-l}, left panel), which moves towards smaller relative lengths as the divider is tilted more ($c$ is larger). This is displayed in the right panel figure \ref{3-W-l}.


We see in figure \ref{3-W-l} that the maximally retained relative relevant information attains a minimum, as a function of $c$. This was also the case for the two-state memory. 

In figure \ref{optimalIfrac} we compare the relative value of captured relevant information at the optimal value of $\ell$ with that captured by the two-state memory. The minimum shifts towards slightly larger $c$, as the number of states increases from two to three. Three states substantially increase how much relevant information can be captured. 
Figure \ref{optimalIfrac} also compares parametrically optimized four- and five-state memories, which we will discuss in the following paragraphs. Note that the above trend continues for larger numbers of states.   

It is intuitive that the optimal coarse graining should reflect the symmetry of the problem, whereby the $m=0$ state is centered around the origin, as is the case in our parametrization, equation (\ref{Irel3}). Asymmetric memories capture less relevant information (which we confirmed numerically).

\subsection{Relevant information for four- and five-state memories} \label{45s}

In the last subsection we saw that for a three-state memory it is always beneficial to choose $\ell < c$. Now we study how much relevant information can be captured by deterministic four- and five-state memories, which we will parameterize similarly. 

The five state memories require two parameters, $\ell$ and $d$. As for the three state memory, $\ell$ parameterizes the center state, which we have called the waste basket, because the error in that state is 1/2. The four state memories requires only one parameter, and because there is no center waste basket, we call this parameter $d$. 

As with the three-state memory, we can allow the length of the parts of the $x$-axis assigned to $m = \pm 1$ to be less than $c$ and it turns out that this is the optimal choice for both, the four- and the five-state memories. 

This means for the four-state memories that out of the three different cases ($d < c$, $d=c$, and $d>c$), $d < c$ is best. 

\begin{table}[h]
\begin{tabular}{ l|c|l } 
$x$ & $m$ & $q_4(m)$ \\ [0.5ex]
\hline
$-1/2 \leqslant x \leqslant -d/2$ & $-2$ & $\epsilon(d,c)$ \\ [0.5ex]
$-d/2 < x \leqslant 0$ & $-1$ & $\hat{\epsilon}(d,c)$ \\ [0.5ex]
$0 < x \leqslant d/2$ & $1$ & $\hat{\epsilon}(d,c)$ \\ [0.5ex]
$d/2 < x \leqslant 1/2$ & $2$ & $\epsilon(d,c)$
\end{tabular}
\caption{Memory assignments and errors for parameterized four-state memories with $d < c$.}
\label{table:4sd}
\end{table}
The parametrization of the assignments and the errors are listed in table \ref{table:4sd} and sketched in figure \ref{fig:4state_err}. In states $m=\pm 2$ we have $q_4(m=-2)= q_4(m=2) =: \epsilon(d,c)$ (equation (\ref{eps})) while $q_4(m=-1)= q_4(m=1) =: \hat{\epsilon}(d,c)$, with
\begin{eqnarray}
\label{eps2}
\hat{\epsilon}(d,c) &=& {1 \over 2} - {d \over 4c}.
\end{eqnarray}
For this memory the relevant information is given by:
\begin{eqnarray} \label{4sl}
I_{\rm rel}^{(4)} &=& (1-d)[\ln(2) - h(\epsilon)] + d\,[\ln(2)-h(\hat{\epsilon})] \notag \\
&=& \ln(2) - (1-d)\, h(\epsilon(d,c) ) - d\,h(\hat{\epsilon}(d,c)).
\end{eqnarray}
For completeness the two suboptimal cases ($d=c$ and $d>c$) are discussed in appendix \ref{AppB}. 

Any symmetric five-state coarse graining again has a waste basket in the center, the length of which we parameterize by $\ell$. The parametrization for the case $d< c$ and $\ell < d$ is shown in table \ref{table:5ld} and figure \ref{fig:5state_err}. Other, suboptimal, cases are discussed for completeness in appendix \ref{AppB}. 

\begin{table}[h]
\begin{tabular}{ l|c|l } 
$x$ & $m$ & $q_5(m)$ \\ [0.5ex]
\hline
$-1/2 \leqslant x \leqslant -d/2$ & $-2$ & $\epsilon(d,c)$ \\ [0.5ex]
$-d/2 < x \leqslant -\ell/2$ & $-1$ & $\hat{\epsilon}(\ell+d,c)$  \\ [0.5ex]
$-\ell/2 < x \leqslant \ell/2$ & $0$ & $1/2$ \\ [0.5ex]
$ \ell/2 < x \leqslant d/2$ & $1$ & $\hat{\epsilon}(\ell+d,c)$ \\ [0.5ex]
$ d/2 < x \leqslant 1/2$ & $2$ & $\epsilon(d,c)$
\end{tabular}
\caption{Memory assignments and errors for parameterized five-state memories with $\ell < d$ and $d < c$.}
\label{table:5ld}
\end{table}

The residual errors for states $m=\pm2$ are \mbox{$q_5(m=\pm2) =: \epsilon(d,c)$,} and for the states $m=\pm1$ we have $q_5(m=\pm1) =: \hat{\epsilon}(\ell+d,c)$.

The relevant information is:
\begin{eqnarray}
I_{\rm rel}^{(5)}
&=& (1-\ell)\ln(2) - (1-d)\,h(\epsilon(d,c)) \notag \\ && - (d-\ell)\, h(\hat{\epsilon}(\ell+d,c)).
\end{eqnarray}
We recover the four-state version (equation (\ref{4sl})) for $\ell \rightarrow 0$. 

\subsection{Extractable work as a function of the tilt}
\begin{figure}[h]
\centering 
\begin{subfigure}{\linewidth}
\centering
\includegraphics[width=.48\linewidth]{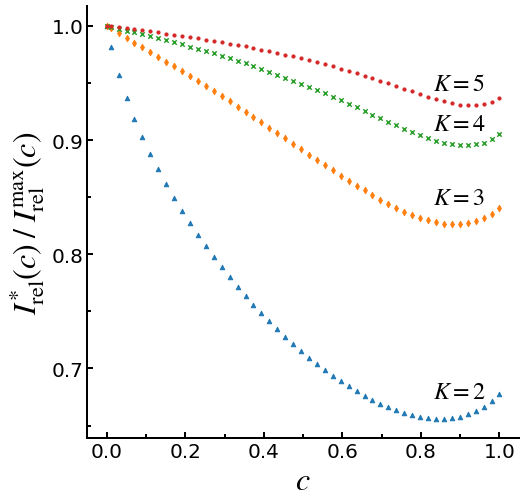}
\includegraphics[width=.48\linewidth]{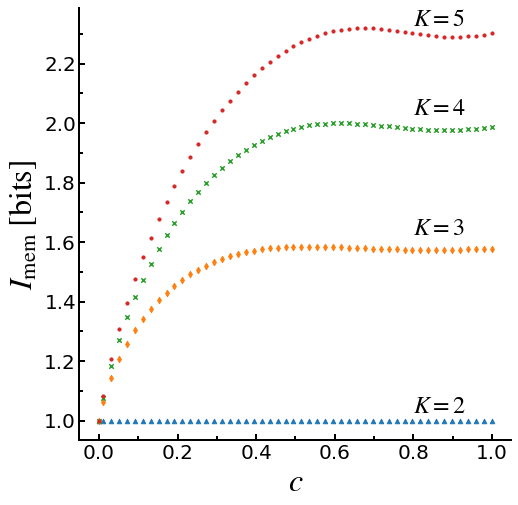} 
\end{subfigure}
\caption{Normalized relevant information (left) for parametrically optimized deterministic partitions of the $x$-axis into $K=2, \dots, 5$ parts, as a function of $c$. Total information (right) captured in memory as a function of $c$, for the parametrically optimized K-partitions that retain maximal relevant information.}
\label{optimalIfrac}
\end{figure}

For different values of the tilt angle, we  numerically found the maximum value of the relevant information, maximized over the parameter(s) of the memory. We denote optimal parameters with $^*$ superscripts, e.g. $\ell^*$ is the parameter value for $\ell$ that maximizes relevant information. The value of the relevant information at the optimal parameters is denoted by $I_{\rm rel}^*(c)$. It changes with the tilt, because the size of the uncertain region, $c$, changes with the tilt. Total available relevant information, $I_{\rm rel}^{\rm max}(c)$, equation (\ref{Imax}), diminishes with increasing divider tilt, i.e. increasing c. We normalize $I_{\rm rel}^*(c)$ by it to investigate how close the relevant information stored in a deterministic $K$-state memory (for $K=2,\dots, 5$) with optimal parameter values is to the maximum attainable relevant information for a given tilt. The results are plotted in figure \ref{optimalIfrac}. 

It is interesting to note that, as we saw for two- and three-states, we find a ``worst" tilt angle for every fixed number of states, in the sense that the parametrically optimized memory captures the least relative amount of relevant information at that angle (and corresponding $c$). As the number of memory states increases, the angle at which this minimum occurs becomes closer to $\theta = \pi/4$, where $c=1$, but it is always larger than $\pi / 4$.

In figure \ref{optimalIfrac} we see that memories with more states can retain a larger fraction of the maximum relevant information than memories with fewer states. The optimal three-state memory always keeps more relevant information than the two-state memory but is in turn outperformed by the best four-state memory in terms of relevant information retained, and so on. As can be expected, returns diminish as the number of states becomes larger: the relative increase in relevant information, gained from adding one more state, becomes smaller. This suggests that further increasing the number of memory states will result in ever diminishing gains in terms of relevant information retained, and thus extractable work. 
The parametrically optimized five-state partition already captures over 93\% of the maximum relevant information for all tilts.

\subsection{Cost of deterministic memories}
As more states get added and the memory becomes more detailed, the cost of running it increases. Running a memory contributes an energetic cost of no less than $kT I_{\rm mem}$ to the engine's overall energy bill \cite{Landauer1961, CB, parrondo2015thermodynamics}. To build intuition, we examine the costs of the deterministic memories we have studied in the last section. 

The amount of information that the memory captures about the data is given by $I_{\rm mem}= I[M,X] = H[M] - H[M|X]$. Since the assignments to memory are deterministic, $H[M|X]=0$, and therefore, $I[M,X] = H[M]$. The entropy, $H[M]$, can readily be evaluated for the different determinisitic memories. 

For two states, it is 
\begin{eqnarray}
I^{(2)}_{\rm mem} &=&  \ln(2).
\end{eqnarray}
The irrelevant information captured by the deterministic two-state memory is thus 
\begin{eqnarray}
I^{(2)}_{\rm irrel} &=& h(c/4), 
\end{eqnarray}
which is zero for $c=0$, the original \LS\ box, and increases monotonically to $c=1$, where the divider is tilted between the two corners of the box, and there are no certain regions.

For three states, we have $p(m=\pm 1) = (1-\ell)/2 $ and $p(m=0) = \ell$, so that:
\begin{eqnarray}
I^{(3)}_{\rm mem} &=&  \ln(2) (1-\ell) + h(\ell).
\end{eqnarray}
This is maximal, with a value of $\ln(3)$, when the waste basket covers the center third of the $x$-axis $\ell=1/3$, and goes to $\ln(2)$, as $\ell$ goes to zero.

Recalling the value of the relevant information, $I_{\rm rel}^{(3)}(\ell)$, equation (\ref{Irel3}), we see that the deterministic three-state memory retains irrelevant information in the amount of 
\begin{eqnarray}
I_{\rm irrel}^{(3)}(\ell) = h(\ell) + (1-\ell)\, h(\epsilon (\ell, c)).
\end{eqnarray}

Four-state assignments have $p(m=\pm 1) = d/2$ and $p(m=\pm 2) = (1-d)/2$, whereby we get
\begin{eqnarray}
I^{(4)}_{\rm mem} &=&  \ln(2) + h(d),  \\
I_{\rm irrel}^{(4)} &=& h(d) + (1-d)\, h(\epsilon(d,c) ) - d\, h(\hat{\epsilon}(d,c)).
\end{eqnarray}

Five-state partitions have probabilities $p(m=0)=\ell$, $p(m=\pm 1) = (d-\ell)/2$ and $p(m=\pm 2) = (1-d)/2$, and thus
\begin{eqnarray}
I^{(5)}_{\rm mem} &=& (1-\ell) \ln(2)  - \ell \ln(\ell) \notag \\&& - (d-\ell) \ln(d-\ell) - (1-d) \ln(1-d), \\
I_{\rm irrel}^{(5)} &=& (1-d)\,[h(\epsilon(d,c))- \ln(1\!-\!d)] - \ell \ln(\ell) \notag \\
&& + (d-\ell)\,[h(\hat{\epsilon}(\ell+d,c)) - \ln(d\!-\!\ell)].
\end{eqnarray}

\begin{figure}[ht]
\centering 
\begin{subfigure}{\linewidth}
\centering
\includegraphics[width=.48\linewidth]{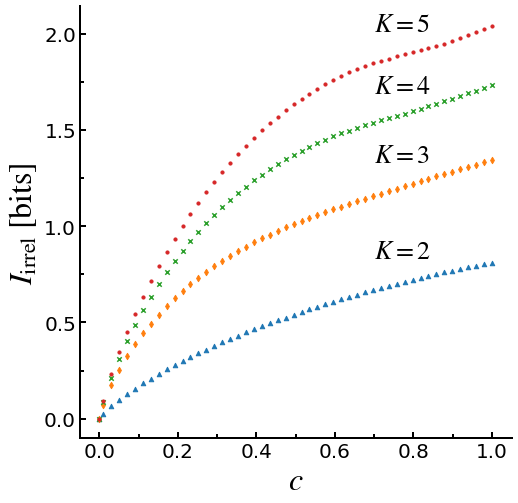}
\includegraphics[width=.48\linewidth]{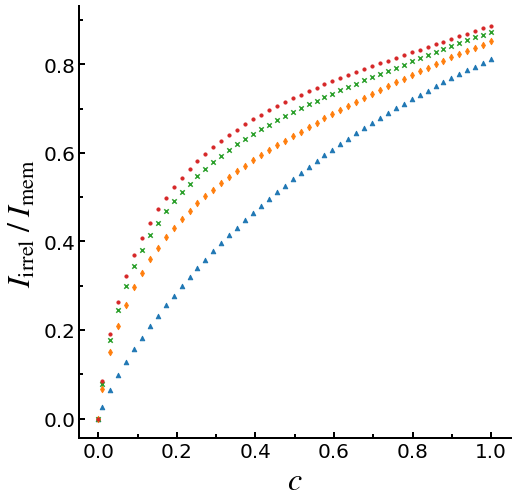} 
\end{subfigure}
\caption{ Total (left) and relative (right) amount of irrelevant information captured in memory as a function of $c$, for the parametrically optimized K-partitions that retain maximal relevant information.}
\label{irrel}
\end{figure}

We compare the memory (figure \ref{optimalIfrac}, right panel) and the irrelevant information (figure \ref{irrel}, left panel) retained by the two-state memory to that retained by the parametrically optimized \mbox{three-,} \mbox{four-,} and five-state memories.
The two-state coarse graining captures a fixed total memory of 1 bit. As the divider becomes more tilted, an increasing amount of that memory encodes irrelevant information. We plot the fraction of irrelevant information kept in memory in figure \ref{irrel} (right panel), and see that for two states it increases monotonically from 0\% to roughly 80\%, as the angle $\theta$ changes between $\pi/2$ and $\pi/4$.

As we recall from figure \ref{optimalIfrac}, coarse grainings with more states can keep a larger fraction of the available {\em relevant} information. This comes at an increased cost in terms of total memory kept, as we see in the right panel of figure \ref{optimalIfrac}. 
When the tilt increases, more resources are utilized, until they are depleted (here the total memory comes to an approximate plateau). However, much of the cost is overhead, capturing increasing amounts of irrelevant information, visible in figure \ref{irrel}, left panel. The relative overhead, $I_{\rm irrel}/I_{\rm mem}$ (plotted in the right panel) is slightly larger, the more states are used.

\section{Optimal data representations balance between cost and gain}
\label{optimalObservers}
The observer part of the engine ought to use a representation of the observed data that allows the information engine to encounter the smallest possible amount of dissipation, on average. Equivalently, it should be constructed such that the engine's average net work output is maximized. 

There is no reason to believe that the best way an observer can summarize the measurements should be given by a {\em deterministic} partition of the observed $x$-values---in general, we ought to optimize over all probabilistic assignments, that is over all conditional probability distributions $p(m|x)$.  

The average heat generated during the memory making process is lower bound by $Q \geqslant kT I[M,X]$, and the average heat absorbed during work extraction is upper bound by $Q' \leqslant kT' I[M,U]$. Therefore, the overall dissipation is no less than $-Q-Q' \geqslant k \left(T I[M,X] - T' I[M,U] \right)$ \cite{CB}.
The total work coming out of the engine for our protocol is given by the amount of work produced in the information-to-work, or work extraction, step, $-W' = kT' I[M,U]$, equation (\ref{Wout}), minus the energetic cost of running the memory, \mbox{$W = kT I[M,X]$}: 
\begin{eqnarray}
- W' - W = k \left(T' I[M,U] - T I[M,X] \right).
\label{Woutengine}
\end{eqnarray}
The amount of heat dissipated with our protocol on average, over a cycle, 
is the negative of equation (\ref{Woutengine}), i.e. $-Q-Q' = W+W'$. 
Equation (\ref{Woutengine}) has to be maximized, over all possible data representations, $p(m|x)$, or equivalently, minimally achievable dissipation is minimized. The data representation that corresponds to the optimum is the one  
we are looking for. It is clear from equation (\ref{Woutengine}) that the solution depends on the choice of the temperature ratio $\tau=T'/T$.
Dividing by $kT$ does not change the optimization problem, we thus find that we need to solve \cite{CB} 
\begin{eqnarray}
\label{IB}
&&\min_{p(m|x)} \Big( I[M,X] - \tau I[M,U]\Big) \notag \\
&&{\rm subject \; to:\;} \sum_m p(m|x) = 1; \; \forall x
\end{eqnarray}
The constraints ensure normalization of the probability distributions. This derivation shows that thermodynamically optimal memories are solutions to the Information Bottleneck \cite{IB} optimization problem.

\subsection{Rate--distortion function}
Written in this form, it is clear that this optimization is a rate-distortion curve calculation \cite{Shannon48, Berger71, IB, StillCru07}. To see that, first recall that $I[M,U] \leqslant I[X,U]$. 
The difference, $I[X,U]-I[M,U]$,  is the information lost because the memory is a {\em summary} of the observable data. This information loss measures how much of the relevant information available from the observable data is discarded in the memory making process. Since $p(u|x) = p(u|x, m)$, the information loss is given by
\begin{eqnarray}
I[X,U] - I[M,U] 
&=& \Big\langle {\cal D}[p(u|x) \| p(u|m)] \Big\rangle_{p(x,m)},
\end{eqnarray}
where ${\cal D}[p(u|x) \| p(u|m)] \equiv \sum_{u} p(u|x) \ln{\left[{p(u|x) \over p(u|m)}\right]} $  is the relative entropy, or Kullback-Leibler divergence, between (a) the probability distribution of the relevant quantity, given knowledge of the observable data, $p(u|x)$, and (b) the probability distribution of the relevant quantity, given knowledge of the memory state, $p(u|m)$. While $p(u|x)$ encodes everything there is to know about $u$, given the observable data, $p(u|m)$ encodes what can be inferred from the memory. Relative entropy measures the mismatch between these two distributions. It quantifies 
the average bit-cost of encoding samples from $p(u|x)$ with a code optimized for $p(u|m)$.
Replacing the optimization in equation (\ref{IB}) by  
\[ \min_{p(m|x)} \Big( I[M,X] + \tau \big\langle {\cal D}[p(u|x) \| p(u|m)] \big\rangle_{p(x,m)} \Big)\]
leaves the solutions unchanged, since $I[U,X]$ does not depend on $p(m|x)$. This shows that the optimization problem is a rate-distortion curve calculation with distortion function ${\cal D}[p(u|x) \| p(u|m)]$. 

\subsection{Information Bottleneck algorithm}
\label{IB-alg-text}
The temperature ratio $\tau = T'/T$ appears in the optimization as the parameter that trades off between how much relevant information the model captures vs how much the model costs. 
  
This ratio parameterizes the family of solutions to equation (\ref{IB}), all of which have to fulfill the following saddle point conditions, a set of self-consistent equations, 
\begin{eqnarray}
p(m|x) &=& {p(m) e^{-\tau {\cal D}[p(u|x) \| p(u|m)]} \over \sum_m  p(m) e^{-\tau {\cal D}[p(u|x) \| p(u|m)]}}, \label{pmx}\\
p(m) &=& \langle p(m|x) \rangle_{p(x)}, \label{pm}\\
p(u|m) &=& {p(u)\over p(m)} \langle p(m|x) \rangle_{p(x|u)}, \label{pum}
\end{eqnarray}
which can be computed numerically by an iterative algorithm. The Information Bottleneck algorithm is similar to the Blahut-Arimoto algorithm \cite{IB,blahut,arimoto}. Numerical calculations are performed using a ``deterministic annealing" procedure, adapted from \cite{rose1998deterministic}. Appendix \ref{AppC} details the numerical implementation. 

\subsection{Characteristics and quality of optimal memories}
\begin{figure}[ht]
\centering 
\includegraphics[width=0.8\linewidth]{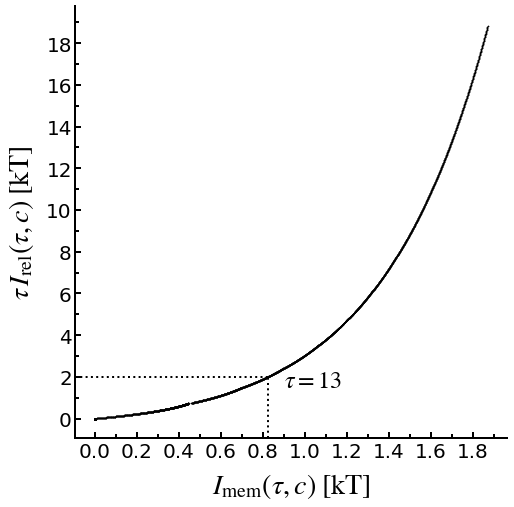}
\caption{Thermodynamic benefit against cost derived from optimal memories (one for each $\tau$), for $c=1$.} 
\label{tau-irel-vs-imem}
\end{figure}

Given a geometry, $c$, there is an optimal memory, for each temperature ratio, $\tau$, fulfilling equations (\ref{pmx})--(\ref{pum}). Let us denote it by $p_{\rm opt}^{\tau}(m|x)$. 
The quality of this family of optimal data representations can be visualized by plotting the work potential retained, that is 
\begin{eqnarray}
kT' I_{\rm rel}(\tau,c) \equiv kT'\, I[M,U] \big|_{p_{\rm opt}^{\tau}(m|x)}, 
\end{eqnarray}
against the energetic cost required to run the corresponding memory, 
\begin{eqnarray}
kT I_{\rm mem}(\tau,c) \equiv  kT\,I[M,X] \big|_{p_{\rm opt}^{\tau}(m|x)}. 
\end{eqnarray}
In units of $kT$, we plot $\tau I_{\rm rel}(\tau,c)$ against $I_{\rm mem}(\tau,c)$, in figure \ref{tau-irel-vs-imem}, 
for the geometry with diagonal tilt, i.e. uncertain region $c=1$.
 
If there is a given constraint on the expectation of how much energy shall at least be extracted from the heat bath, and turned into work, during the information-to-work step of the engine, then the memories that form this curve allow us to determine the smallest value of the temperature ratio 
at which this is possible, and they  furthermore tell us the minimum cost that the observer has to encounter, in order to enable this amount of work extraction, at this lowest possible temperature ratio. We visualize this in figure \ref{tau-irel-vs-imem} for an example by drawing a horizontal line at $\tau I_{\rm rel}(\tau,c)=2$, reading off the value $\tau=13$ at the solution, and plotting it next to the curve, and then reading off the cost of the solution at this point (vertical line), $I_{\rm mem}(\tau,c)=0.82$. 

\begin{figure}[ht]
\centering 
\includegraphics[width=\linewidth]{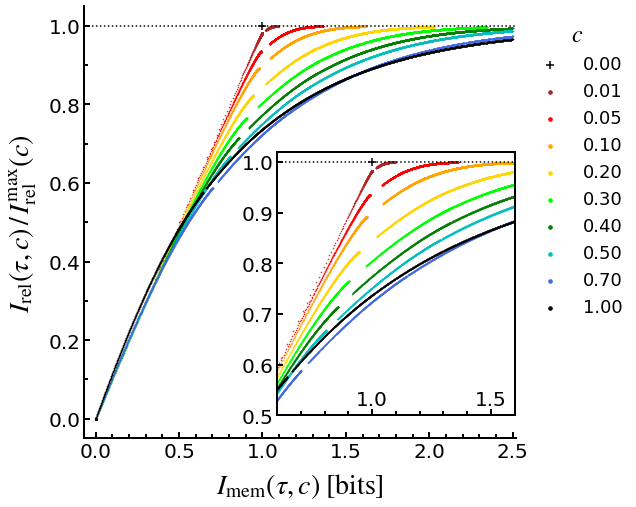}
\caption{ Normalized extractable work, against total information kept in memory, 
evaluated at the optimal memories (one for each $\tau$), for different values of $c$. Inset magnifies the most interesting region.}
\label{IBcurves}
\end{figure}

To compare curves like the one in figure \ref{tau-irel-vs-imem} for {\em different} geometries, i.e. different $c$, we divide the 
extractable work at the optimal data representation in units of $kT$, that is $\tau I_{\rm rel}(\tau,c)$, by the maximum, given the geometry (also in units of $kT$), i.e. $\tau I_{\rm rel}^{\rm max}(c)$, equation (\ref{Imax}).
The ratio, $I_{\rm rel}(\tau,c)/I_{\rm rel}^{\rm max}(c)$, is plotted in figure 
\ref{IBcurves} against the minimum work cost in units of $kT$, $I_{\rm mem}(\tau,c)$. For visual ease, we plot the information in bits, i.e. $I_{\rm mem}(\tau,c)/\ln(2)$.

Geometries with only slight tilts (small c) are dominated by large certain regions. Their optimal curves are steeper, with larger curvature, and approach \mbox{$I_{\rm rel}(\tau,c)/I_{\rm rel}^{\rm max}(c)=1$ }at smaller memory costs. 
In that sense, they are more compressible (the notion of source compressibility is adapted from \cite{StillCru07}, see appendix \ref{compress}). 

Each of the curves can be used to read of the best possible code if capacity constraints are imposed on the memory.
We see in figure \ref{IBcurves} that the relative amount of relevant information captured by memory encodings that encounter the same cost decreases as the tilt increases. 

\begin{figure}[ht]
\centering 
\includegraphics[width=0.9\linewidth]{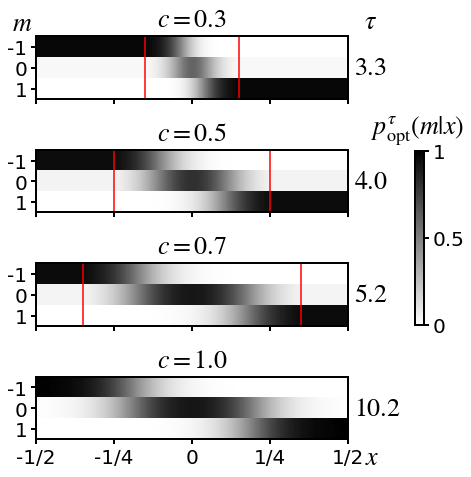}
\caption{ Memory assignments of optimal memories with $I_{\rm mem}(\tau,c) = 1$ bit, for different geometries, $c$. Corresponding $\tau$ displayed to right. Red lines mark $\pm c/2$, where the divider touches the top/bottom of the box.}
\label{1bit}
\end{figure}

As we know, the optimal solution for the original \LS\ engine geometry, without tilt, is a deterministic coarse graining into two memory states. This retains 1 bit, all of which is relevant information. This solution is (trivially) found by the numerical algorithm for all values $\tau \geqslant 1$, and the point is plotted as a cross in figure \ref{IBcurves}. In figure \ref{1bit}, we compare optimal codes that cost exactly one bit of information, for four different geometries, by visualizing their assignments. 
The one bit is distributed between three memory states. The third memory state gets utilized more as the tilt increases. 

Note that for the original \LS\ engine for $\tau=1$ a second solution exists producing the same amount of (zero) dissipation. The second solution corresponds to doing nothing, i.e. the point (0,0) in the information plane of figure \ref{IBcurves}, for which we did not plot the cross symbol, because it coincides with the solution at $\tau=1$ for all tilt angles, and would thus obscure the plot. The original \LS\ engine is trivially fully compressible, because it is fully observable, meaning that all relevant quantities can be measured. 

For \LS's engine, these two are the only optimal solutions, whereby the information plane is reduced to two points, with a gap in between. Similarly, we find visible gaps at phase transitions \cite{wu2020phase} to increasing numbers of memory states, in the information curves for different tilt angles (figure \ref{IBcurves}). For geometries with $c \leqslant 0.2$, one-bit memories fall into the phase transition gap (which is why we display $c \geqslant 0.3$ in figure \ref{1bit}). 

For each individual geometry, $c$ is fixed, as is $I^{\rm max}_{\rm rel}(c)$, equation (\ref{Imax}), and we can plot retained relevant information at the optimal solution of equation (\ref{IB}), $I_{\rm rel}(\tau)$, as a function of total memory, $I_{\rm mem}(\tau)$. The optimal memories draw out a monotonic, concave curve in the information plane \cite{IB}, akin to a rate-distortion curve \cite{Berger71}. The curve delineates feasible from infeasible region: no codes exist above the curve, and the codes that lay below the curve are suboptimal. The local slope of the curve is given by $ T/T'= {1 / \tau}$, which is $1$ at the origin and approaches zero asymptotically, as all of the relevant information is captured and the memory's size increases. 

\begin{figure}[ht]
\centering 
\includegraphics[width=0.9\linewidth]{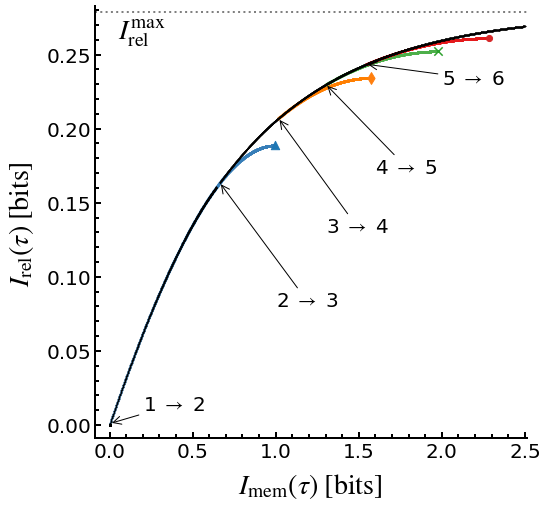}
\caption{ Relevant information against total information kept in memory, evaluated at the optimal memories (one for each $\tau$, black curve) displayed for $c=1$. Colorful curves are evaluated for memories that are optimal under the additional constraint of allowing no more than K states. $K=2$ (blue), $K=3$ (orange), $K=4$ (green), $K=5$ (red).}
\label{Infoplane-theta45}
\end{figure}

We show the optimal memories in the information plane for the least compressible geometry, $c=1$, 
(where the divider is tilted to $\pi/4$, touching the corners of the box) in figure \ref{Infoplane-theta45}. The upper bound on relevant information (displayed on the $y$-axis) is $I_{\rm rel}^{\rm max}(c=1) = \ln(2) - 1/2$ nats, or $1-1/\ln(4) \approx 0.28$ bits, which is shown as a dotted line. On the $x$-axis, the upper bound on how much total information can be retained about a position measurement should follow from the measurement error, which would be given by instrumentation. However, numerically, we approximate the continuous $x$ positions in the box by 10000 discrete values. The maximum amount of information that can be captured in memory is then set to a little more than 13 bits. We cut off at 2.5 bits, where the optimal code already captures 96.6\% of the available relevant information. 

The solution for $\tau=1$ assigns all $x$-values to one memory state. This captures no information, and does not allow us to exploit the partially observable \LS\ engine---when $T'=T$, then the best solution is to do nothing. 

As the temperature ratio $\tau = T'/T$ increases, this solution becomes suboptimal, and a phase transition occurs to a solution that uses two memory states. The assignments from observed $x$ measurements to the memory states are not deterministic at the phase transition. As $\tau$ is increased, they become increasingly less probabilistic, in the sense that $H[M|X]$ becomes smaller. As one traces out the optimal curve, phase transitions occur to memories with increasing numbers of memory states. In figure \ref{Infoplane-theta45}, we point to those with arrows labeled $K \rightarrow K+1$, indicating the transition from $K$ to $K+1$ memory states for the first five transitions. (It can be expected from the geometry of our problem that transitions occur to $K+1$ states, but this is not necessarily true in general \cite{IB,wu2020phase}.)

\begin{figure}[ht]
\centering 
\includegraphics[width=0.9\linewidth]{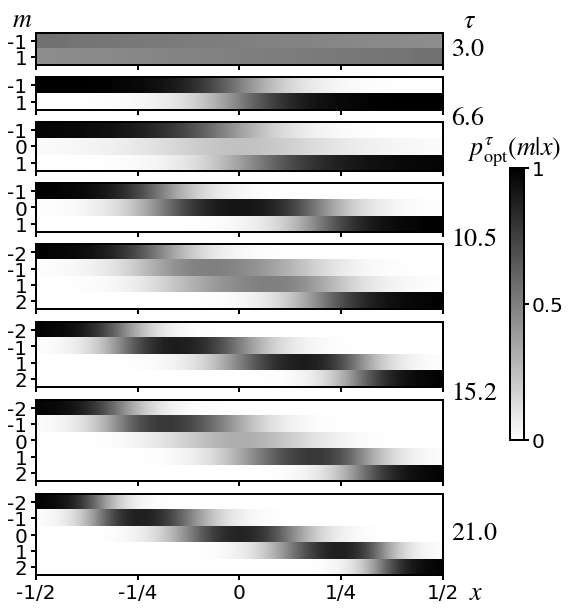}
\caption{Optimal, probabilistic, assignments from observables to memory states before and after the phase transitions. Displayed is the optimal probability of using memory state $m$, given that $x$ is observed: $p_{\rm opt}^{\tau}(m|x)$. From top to bottom, the first assignment after a phase transition and the last one before the next phase transition are displayed for increasing numbers of states, starting with $K=2$. Geometry with $\pi/4$ tilt, i.e., $c=1$. Approximate values of $\tau$ at the phase transitions are displayed to the right.}
\label{transitions_c1}
\end{figure}

To obtain a better intuition for what the optimal data representations look like, we visualize them at the phase transitions in figure \ref{transitions_c1}. The first panel shows the first optimal two-state solution right after the phase transition. We see that it contains very little information, as the assignments are close to uniform. This changes as $\tau$ increases. The second panel shows the optimal two-state solution right before the next phase transition. The third panel shows the optimal three-state solution right after that phase transition, and so forth. This visualizes how the assignments become increasingly deterministic in between phase transitions, but that optimal memories are {\em not} deterministic assignments. Instead, optimal data representations assign observations of $x$ positions probabilistically to more than one memory state.

One can force the number of states in the numerical calculation to be equal to two, while increasing $\tau$. The solutions to the self-consistent equations obtained under this constraint then branch off from the optimal curve in the information plane, at the point where the next phase transition occurs (blue points in figure \ref{Infoplane-theta45}). 
The constrained two-state solution will eventually become deterministic, as $\tau$ is increased. At that point, 1 bit of information is kept in memory, less than 0.19 bits of which are useful, in the sense of being relevant with respect to allowing the observer to extract work from the system. This point in the information plane is plotted as a triangle in figure \ref{Infoplane-theta45}. The solution coincides with the deterministic two-state assignment we studied in section \ref{WE-mems}, see equation (\ref{Irel2}). 

We trace out all curves with fixed $K\leqslant 5$ states in figure \ref{Infoplane-theta45}. Their endpoints coincide with the deterministic, parametrically optimized, partitions we studied in \mbox{section \ref{WE-mems}}. They are plotted using the same symbols as in figure \ref{optimalIfrac}. Compare the richness of the full optimization problem and its solutions to the naive analysis of section \ref{WE-mems} in the information plane: the latter produces only the end-points of the constrained, suboptimal curves, whereas solving equation (\ref{IB}) provides us with the entire family of optimal solutions, one for each $\tau$. 

The choice of memory thus depends on the cost-benefit trade-off parameter, $\tau$. For a given geometry, there is an observer memory for each value of $\tau$, which enables minimization of the dissipated heat per cycle. Using a costlier memory becomes worth it only if the temperature ratio, $\tau$, is increased to a point where the added benefit outweighs the additional cost. 

If the observer faces constraints, say metabolic constraints, limiting maximally allowable costs to $I_{\rm mem}^{\rm constr}$, then there is a maximum thermodynamic benefit that can be achieved at a minimum $\tau^*$. At this temperature ratio, the optimal memory carries $I_{\rm mem}(\tau^*) = I_{\rm mem}^{\rm constr}$. 
The magnitude of the corresponding maximum thermodynamic benefit can be read off from the curve in the information plane, as it is determined by the relevant information, $I_{\rm rel}(\tau^*)$.
Conversely, if a fixed minimum thermodynamic gain is required, the minimum cost to achieve it, at the minimal possible temperature ratio, can be read off. 

\begin{figure}[ht]
\centering
\begin{subfigure}{\linewidth}
    \includegraphics[width=0.85\linewidth]{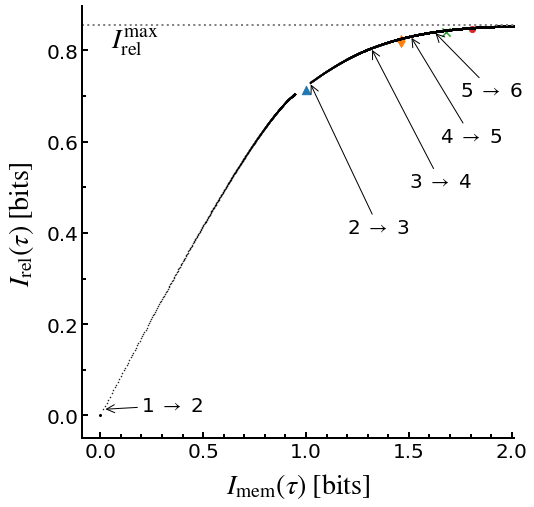}
\end{subfigure}
\begin{subfigure}{\linewidth}
   \includegraphics[width=0.85\linewidth]{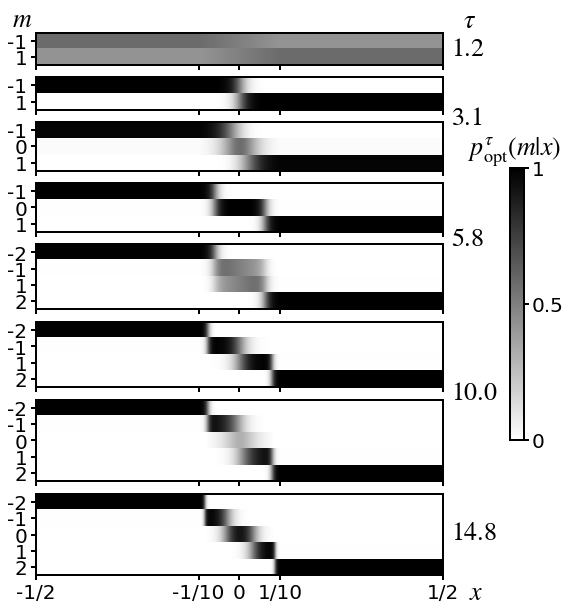}
\end{subfigure}
\caption{ Top: optimal memories evaluated in the information plane for $c=0.2$. Bottom: corresponding memory assignments before and after phase transitions.}
\label{c02}
\end{figure}

Gaps in the curve due to phase transitions are obscured in figure \ref{Infoplane-theta45} by the colored curves that trace out the solutions with constraints \mbox{$K\in \{2, \dots, 5\}$}. In figure \ref{c02} we plot the information plane for $c=0.2$, together with the corresponding optimal assignments before and after phase transitions. 
As before, arrows point to phase transitions from $K$ to $K+1$ memory states on the curve, but we now plot only the end points of the constrained, suboptimal $K$ state memories, not the entire constrained curves, and the gaps at $1\rightarrow 2$ and $2\rightarrow 3$ are therefore visible. Looking back at the comparison plot in figure \ref{IBcurves}, we see that the $2\rightarrow 3$ gap is larger for $c=0.2$, compared to $c=1$.

Compare the assignments in figure \ref{c02} to the assignments for the stronger tilt, figure \ref{transitions_c1}. Because the certain regions start to dominate for $c=0.2$, assignments in those regions, towards the walls of the box, are largely almost deterministic before the phase transition from 2 to 3 states, and remain so thereafter. The probability weight of the new state ($m=0$) is thus more concentrated towards the middle, and the assignments after the phase transition are less fuzzy compared to the geometry with the diagonal tilt and no certain regions, as can be seen in figure \ref{transitions_c1}, third panel from the top. 

The three-state solution right after the phase transition is thus relatively more costly for $c=0.2$ than for $c=1$, but it also contains relatively more relevant information. This explains why the gap at the phase transition is larger.

For the geometry with $c=0.2$, the deterministic two-state memory falls just below the gap (triangle, figure \ref{c02}). The difference between optimal memories and the deterministic, parametrically optimized, $K$-state memories of section \ref{WE-mems} is less pronounced for $c=0.2$, compared to $c=1$, because the certain regions dominate more. These effects are even stronger for smaller tilts, as the dominance of the certain regions increases. The relative contribution to $I_{\rm rel}^{\rm max}(c)$ from the certain regions is $I_{\rm rel}^{\rm cer} / I_{\rm rel}^{\rm max} = (1 - c) / [1 - c/ \ln(4)]$, plotted  as a function of their combined size, $1-c$, in figure \ref{rel-cont-cert} (detailed calculation in appendix \ref{App:rel_certain}).
\begin{figure}[ht]
\includegraphics[width=0.65\linewidth]{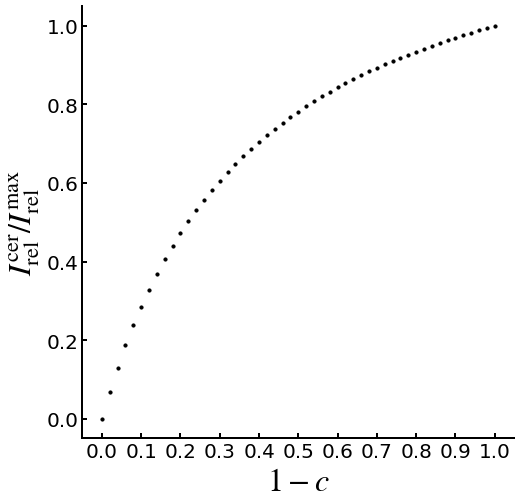}
\caption{Relative contribution to relevant information from the certain regions, as a function of the combined size of the certain regions, $1-c$.}
\label{rel-cont-cert}
\end{figure} 

\subsection{Phase transitions for different geometries}
\label{phase-trans}
At a phase transition, the total information captured in memory increases disproportionally more than the amount of relevant information retained. This is a consequence of minimizing the lower bound on dissipation, because when the temperature ratio changes, the relative contributions of cost versus potential gain are weighted differently. The discontinuities in figures \ref{tau-irel-vs-imem}, \ref{IBcurves}, \ref{Infoplane-theta45}, \ref{c02}, and \ref{eff}--\ref{Irrel/Rel} are due to this effect.

\begin{figure}[ht]
\centering 
\includegraphics[width=0.9\linewidth]{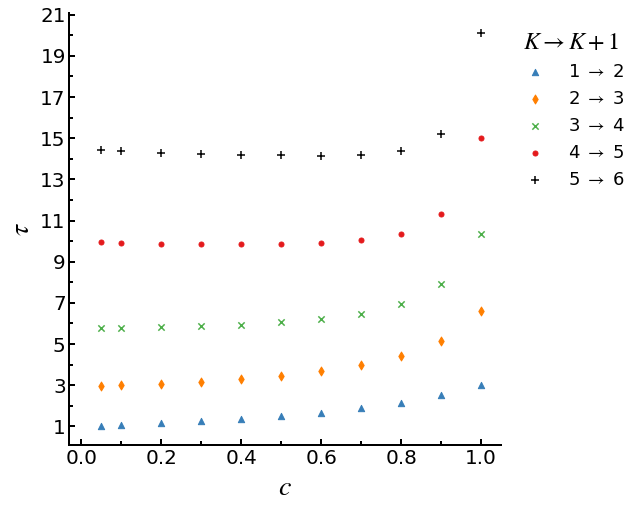}
\caption{ Value of $\tau$ at phase transitions, where the number of memory states changes from $K$ to $K+1$. }
\label{tau-crit}
\end{figure}

Phase transitions occur at larger $\tau$, as the tilt increases, and with it the size of the uncertain region, $c$. The trend is highly non-linear, as can be seen in figure \ref{tau-crit}. A significant region is almost flat, before there is a strong increase. The nonlinearity is more pronounced for larger $K$.

\section{Constructing minimally dissipative memories}
\label{Sec:opt_mem}
The preceding analysis showed that minimally dissipative memories require {\em probabilistic} partitions, instead of the more familiar deterministic coarse grainings. In fact, the naive approach of looking for parametrically optimized deterministic coarse grainings, could never discover minimally dissipative memories (cf section \ref{WE-mems}, and appendix \ref{Det-mindiss}). 

Probabilistic coarse grainings are not common in physics, and are not intrinsically intuitive. 
To picture how to physically construct a probabilistic memory, let us consider the example of implementing the memory with another one particle gas in a container with unit height, width, and depth, at temperature $T$. Recall that there is one optimal memory for each value of $\tau$. This optimal memory is characterized by the solutions to optimization equation (\ref{IB}), that is, the probabilities $p_{\rm opt}^{\tau}(m)$ and $p_{\rm opt}^{\tau}(m|x)$ which fulfill equations (\ref{pmx})-(\ref{pum}).

\begin{figure}[ht]
\centering 
\includegraphics[width=\linewidth]{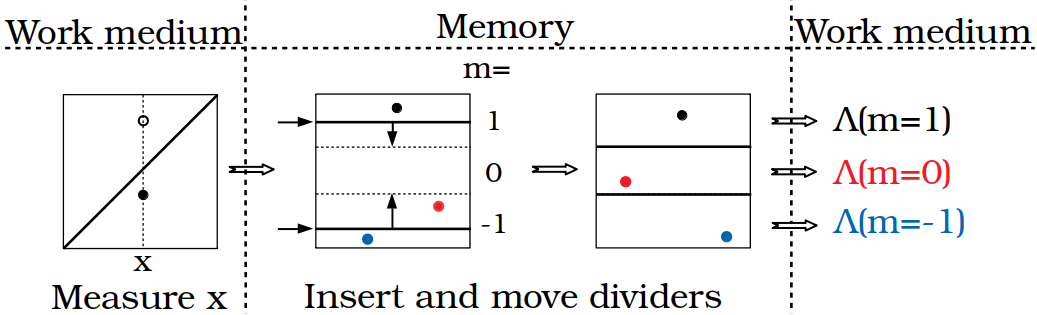}
\caption{Mechanism to implement optimal memories, as described in the text. Sketched here is an example with $c=1$, $\tau=7.76$, $x=0.02$. The three different cases that can occur are visualized in color, for particle location and corresponding work extraction protocol applied to work medium (section \ref{workextraction}): $\Lambda (m=\pm 1)$ partition moves left/right (black/blue), leaving optimized volume 
equal to $p(u\neq m |m)$; 
$\Lambda(m=0)$ partition does not move.} 
\label{memory-making-sketch}
\end{figure}

Arrange the memory read-out such that each state, $m \in \{1,\dots,K\}$, is determined by the $y$ position of the particle in the memory box ($y$ is chosen to avoid linguistic and visual confusion with the work medium, of which the $x$ position is measured; this choice can be made without loss of generality). For any given measurement, $x$, run the following protocol on the memory: 
\begin{enumerate}[label=(\alph*)]
\item To ensure that the particle resides in each memory state with probability $p_{\rm opt}^{\tau}(m|x)$, we want to create initial volumes of size $p_{\rm opt}^{\tau}(m|x)$. This is done by inserting an array of $K-1$ dividers, indexed by $a = 1,\dots, K-1$, into the memory cube at positions $\sum_{m \leqslant a} p_{\rm opt}^{\tau}(m|x)$ (see figure \ref{memory-making-sketch}). The first divider ($a=1$) is inserted at position $p_{\rm opt}^{\tau}(m=1|x)$, the last divider at position $1-p_{\rm opt}^{\tau}(m=K|x)$. 

\item To be able to read out the memory without knowledge of $x$, move the dividers to positions $\sum_{m \leqslant a} p_{\rm opt}^{\tau}(m)$, $\forall a$. These final positions do not depend on $x$, and the value of $m$ can henceforth be read off without knowledge of $x$. 
\end{enumerate}

As a result, of this preparation, state $m$ will occur with probability $p_{\rm opt}^{\tau}(m|x)$, when $x$ is given, and with probability $p_{\rm opt}^{\tau}(m$) on average (over all $x$). The work extraction protocol (labeled $\Lambda(m)$ in figure \ref{memory-making-sketch}, described in section \ref{workextraction}) is then chosen according to which memory state, $m$, is read out. 

The average work done on the memory in this process is precisely $kT I_{\rm opt}^{\tau}[M,X]$, in the quasi-static limit: with probability $p_{\rm opt}^{\tau}(m|x)$, the volume will be changed from $V_i=p_{\rm opt}^{\tau}(m|x)$, to $V_f = p_{\rm opt}^{\tau}(m)$. The resulting work is $W(m,x) = kT \ln[ p_{\rm opt}^{\tau}(m|x) / p_{\rm opt}^{\tau}(m) ]$. The average work per observation is then $W(x) = \sum_m p_{\rm opt}^{\tau}(m|x) W(m,x)$, and the overall average $\langle W(x) \rangle_{p(x)} = kT I_{\rm opt}^{\tau}[M,X]$.

Since it is not immediately intuitive what the optimal divider arrays for the memory making protocols should look like,
we further visualize them in figure \ref{pistons-codebook}, for selected $x$ measurement values located in the left half of the work medium box, for each of the optimal solutions displayed in figure \ref{transitions_c1} (the geometry with diagonal tilt, $c=1$). 
Due to symmetry, plotting $x$ values in the other half would be redundant. Displayed on the left of figure \ref{pistons-codebook} is the $x$ position at the edge of the container; moving towards the right, $x$ positions increasingly closer to the center, $x=0$, are displayed. 

In each of the subplots for different $x$ positions, the probabilities $p_{\rm opt}^{\tau}(m)$ are displayed along on the $x$-axis of the plot, as the widths assigned to each value of $m$. For visual ease, the $x$-axis is labeled with the values of $m$ only in the top row of each two rows with the same number of memory states, $K$, and solid lines divide between solutions with different $K$.
In each subplot, a solid line is plotted across the width of each state, at $p_{\rm opt}^{\tau}(m|x)$ on the $y$-axis. Grey shades are added to guide the eye.
The heights of the white areas thus correspond to the initial volumes delineated by the divider array in the memory making process (figure \ref{memory-making-sketch}), while the widths correspond to the final volumes.
The last subplot (far right) displays the probabilities $p(u|m)$, which give the individual memory states their meaning via the inference they imply. 
(Recall that $u=\{-1,1\}$, when the $\{$right, left$\}$ side is empty).

For example, in the second line, the optimal two state memory right before the phase transition is plotted. Here, the state $m=-1$ implies a larger probability that the right side is empty, corresponding to a lower $p(u=1|m)$, while $m=1$ implies that the left is empty with higher probability. No compression of the particle in the memory box occurs if the particle is observed in the center ($x=0$): then the divider is inserted in the middle and left there. Visually, this corresponds to two white squares in the subplot. In contrast, the white rectangle in the subplot on the left ($x=-1/2$) indicates that if the particle falls into the larger initial volume, then compression will occur: the divider is inserted close to the edge of the container and moved to the middle. This procedure will most likely result in the memory state $m=-1$, because the particle will reside in the larger volume with higher probability. In the unlikely event that the particle falls into the smaller initial volume, expansion occurs, negative work is done on the memory (meaning that work is derived from it), and we end up with an inaccurate inference, resulting most likely in loss during the work extraction protocol performed on the work medium.

While certain memory states dominate for certain $x$ values, it is never the case that there is no residual volume in any of the other memory states (as there would be for a deterministic coarse graining). As a result, for odd numbers of states, when $x=0$ is measured, there is a chance that the chosen protocol, $\Lambda(m)$, will move the piston in the work medium. This is not immediately expected or intuitive, but it makes sense when {\it both} the minimal cost and the maximal possible gain are taken into account.

The mapping from observations to memory visualized here is highly nontrivial, and could not easily have been discovered without the use of the iterative algorithm (section \ref{IB-alg-text}). Due to space limitations, only five $x$ positions are displayed in figure \ref{pistons-codebook}, but remember that they are drawn, in principle, out of a continuum. A movie, changing with $\tau$, to greater resolution
in $x$ is supplied online \footnote{See supplementary material at [\url{https://stacks.iop.org/NJP/24/073031/mmedia}] for a movie of the visualization of the physical codebook displayed in figure \ref{pistons-codebook}, to higher precision in $x$ and $\tau$.}.

\begin{widetext}
\begin{minipage}{\linewidth}
\begin{figure}[H]
\centering 
\includegraphics[width=\linewidth]{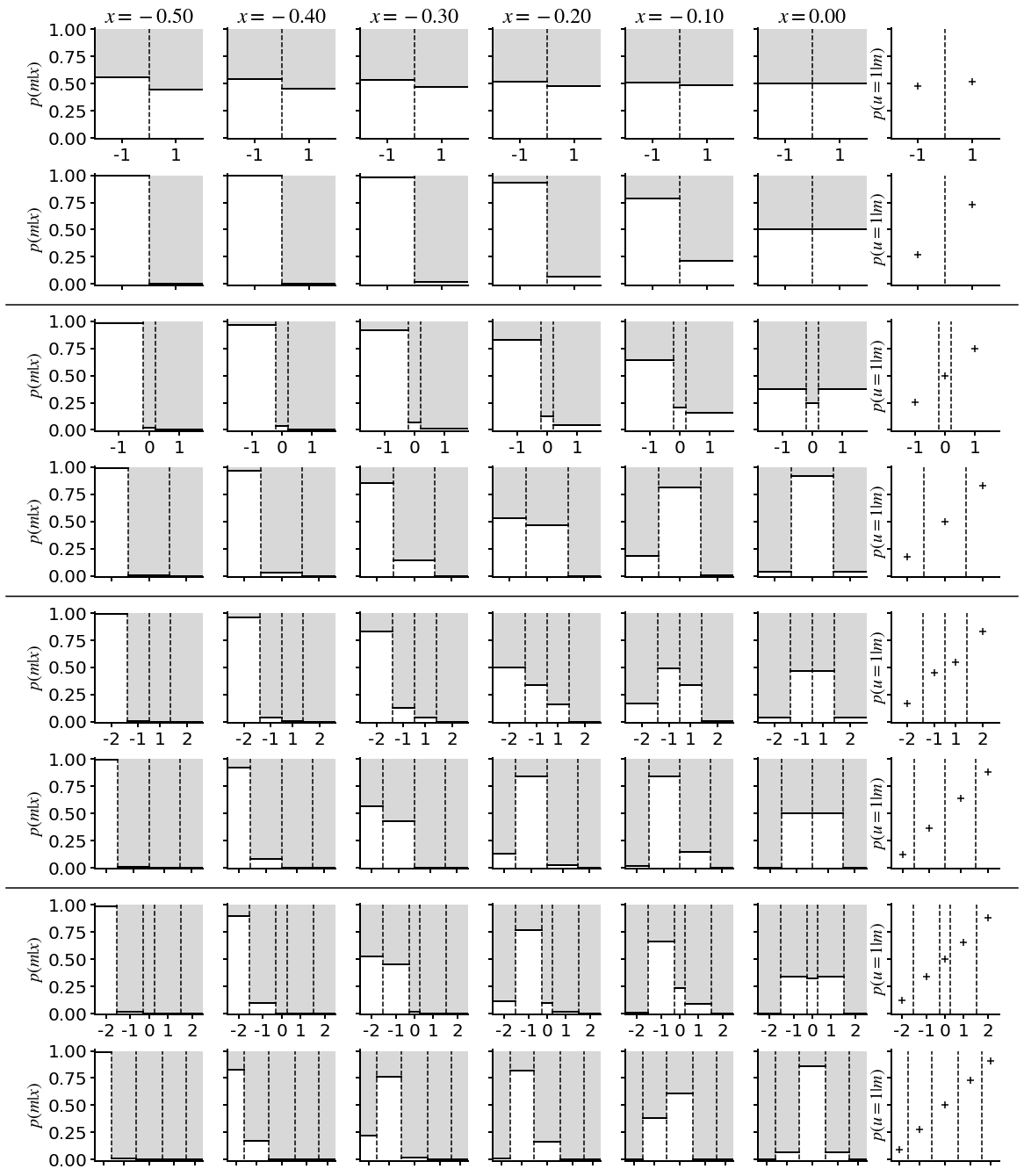}
\caption{Visualization of the method for making optimal memories corresponding to those displayed in figure \ref{transitions_c1}.}
\label{pistons-codebook}
\end{figure}
\end{minipage}
\end{widetext}

\vspace*{3cm}

\section{Thermodynamic efficiency of engines with optimal memories}
\label{eta}
Thermodynamic efficiency, defined as the total work output divided by the heat absorbed at the higher temperature, $\eta = (-W'-W)/Q'$, is a classical performance measure for engines. Importantly, no engine can surpass the Carnot efficiency, $\eta_C = 1 - T/T'$. For generalized information engines, we have \cite{CB}: 
\begin{eqnarray}
\label{efficiency}
\eta &=& 1 - {T \over T' }{I_{\rm mem} \over I_{\rm rel}}.
\end{eqnarray}
Memories that contain relatively more relevant information, i.e. larger $I_{\rm rel} /I_{\rm mem}$, result in higher thermodynamic efficiency at the same value of $\tau= T'/T$.

\begin{figure}[ht]
\centering 
\includegraphics[width=0.95\linewidth]{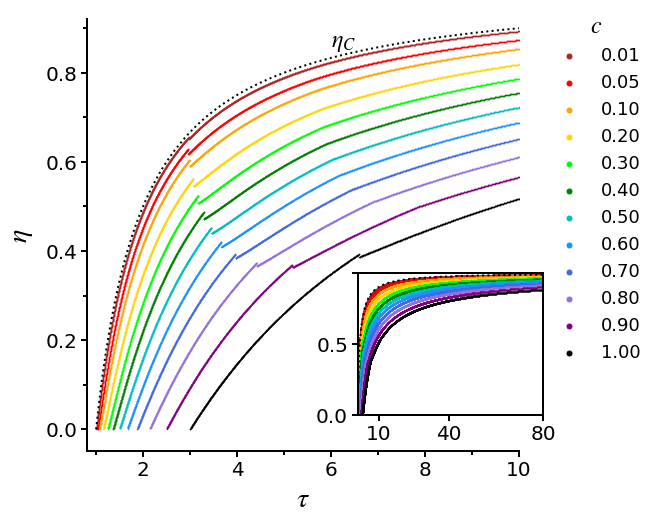}
\caption{ Efficiency achieved with optimal memories as a function of $\tau$, plotted for different tilt angles. Dotted line: Carnot efficiency, $\eta_C$.}
\label{eff}
\end{figure}

In the previous section, we showed that for each value of $\tau$, there is a memory that maximizes net engine work output and minimizes dissipation. These optimal memories have the feature that no other memory that costs the same can yield more, i.e. out of all memories with the same thermodynamic cost of $kT I_{\rm mem}$, the optimal memory enables the largest gain, $kT' I_{\rm rel}$, for a fixed temperature ratio $\tau$. 
Under the constraint of partial observability, Carnot efficiency is not achievable, 
because a solution with $I_{\rm rel} = I_{\rm mem}$ is infeasible for $\tau>1$ when the divider is tilted (see figure \ref{IBcurves}).
The efficiency that can be achieved with the minimally dissipative memories is plotted in figure \ref{eff}, where it is compared to the Carnot efficiency. 

For geometries with smaller tilts, optimal memories result in efficiencies closer to $\eta_C$. Their behavior is dominated largely by the temperature ratio, but as the tilt increases (larger $c$) the influence of information processing inefficiency becomes 
stronger. The curves have discontinuities at phase transitions to more memory states.

It is clear from equation (\ref{efficiency}) that thermodynamic efficiency, as defined for heat engines, extends naturally to generalized, partially observable information engines. Standard information engines are a subclass of generalized information engines, with $\tau=1$, and full observability, hence trivially $I_{\rm rel} = I_{\rm mem}$. In the quasi-static limit a standard information engine is able to recover the cost of running the memory, and hence can reach zero efficiency. Away from thermodynamic equilibrium, the average thermodynamic efficiency of stochastic standard information engines is non-positive.
These facts serve as a useful reminder of the inherent limitations of 
information engines. 

The ratio of average work extracted per $kT I_{\rm mem}$ has been used as an alternative efficiency measure for standard information engines (e.g. \cite{proesmans2015efficiency}, and references therein). In the quasi-static limit, this measures the ratio of thermodynamic gain over thermodynamic cost, given by $\tau I_{\rm rel} /I_{\rm mem}$ for generalized information engines, and given by 1 for standard information engines. 
Since quasi-static operation provides a bound for engines operating away from equilibrium, the average of this measure does not exceed 1 for stochastic engines \cite{proesmans2015efficiency}. 
It is important to remember that using this measure in place of $\eta$ (equation (\ref{efficiency}))
distracts from the fact that
standard information engines cannot have positive thermodynamic efficiency. At best they can  recuperate the thermodynamic cost of operating the memory they need to function. 

\begin{figure}[ht]
\centering 
\includegraphics[width=0.95\linewidth]{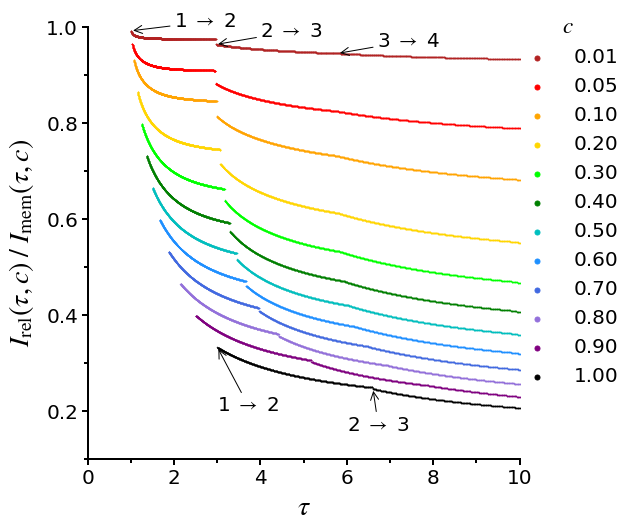}
\caption{ Relevant information per unit of total information kept in memory as a function of $\tau$ for different divider tilts (different $c$).}
\label{rel_mem_ratio}
\end{figure}

Figure \ref{rel_mem_ratio} shows the relevant information per bit of memory, ${I_{\rm rel}(\tau,c) / I_{\rm mem}(\tau,c)}$, evaluated at the optimal memories, as a function of $\tau$, plotted for different geometries (different values of $c$). Phase transitions are marked on the curves for the diagonal tilt (lowest curve, $c=1$), and for the slightest tilt displayed (top curve, $c=0.01$). As $c$ decreases, curves asymptote to 1, which is reached by the original \LS\ engine.

While $I_{\rm rel} /I_{\rm mem}$ gives interesting insights into the structure of the process at hand, using it as an information processing efficiency measure would be misleading, 
as it would suggest that the fuzziest memories (points at the left edges of the curves in figure \ref{rel_mem_ratio}), which keep almost no information at all (see figures \ref{transitions_c1} and \ref{c02}), achieve the highest efficiency, and are therefore ``best" in that sense. It is crucial to remember that the ``goodness" of a memory can only be judged under a constraint: of two memories with the same cost ($I_{\rm mem}$), the one that captures more useful, relevant information (larger $I_{\rm rel}$) is better. Of two memories that contain the same amount of relevant information, $I_{\rm rel}$, the one that is more concise (smaller $I_{\rm mem}$) is better. For a given cost-benefit trade-off (fixed $\tau$) there is a best memory, but which memory is best overall cannot be decided. 

Is the memory that does not summarize anything at all, i.e. with continuous $m=x$, the best memory, because it captures all of the available relevant information? But it does not summarize anything. Is a two-state memory not better, because it achieves substantial compression? These questions have no answer, all we can say is that using a more detailed memory is not worth it, until the temperature ratio $\tau$ reaches a certain value. For example, in the geometry with $c=0.2$, using a two state memory is not worth it, until $\tau$ reaches roughly 1.2, see figure \ref{c02}.

\begin{figure}[h]
\centering 
\includegraphics[width=0.95\linewidth]{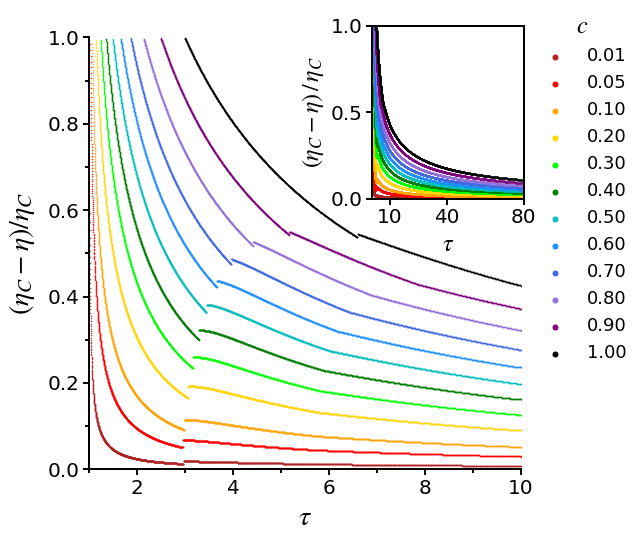}
\caption{ Fractional reduction from Carnot Efficiency $\eta_C$ as a function of $\tau$ for different divider tilts (different $c$).}
\label{Eta}
\end{figure}

It is instructive to note that, in general, the reduction from Carnot efficiency depends on the ratio of irrelevant to relevant information kept in memory
\begin{eqnarray}
\eta_C - \eta = {T \over T' }{I_{\rm irrel} \over I_{\rm rel}}.
\end{eqnarray} 
Relative reduction is plotted as a function of $\tau=T'/T$ in figure \ref{Eta} 
for the optimal memories (that solve the optimization problem, equation (\ref{IB}), and thereby result in minimal dissipation). 
Comparing curves for geometries with a smaller tilt to those with a larger tilt, we see that for small tilts, the Carnot efficiency is approached at smaller $\tau$ values, because these geometries result in more compressible systems (in the sense of data compression). For example, when $\tau=2$, the geometry with the slightest tilt ($c = 0.01$) can operate close to Carnot efficiency, with a relative reduction of only roughly $2.4\%$. At the same value, $\tau=2$, the geometry with the diagonal tilt uses one memory state and thus cannot be used for work extraction. For this geometry, the relative reduction from Carnot efficiency is significantly lowered only at much larger $\tau$ values (inset of figure \ref{Eta}), and even at $\tau=80$, the relative reduction is still roughly $10.6 \%$.

\begin{figure}[h]
\centering 
\includegraphics[width=0.95\linewidth]{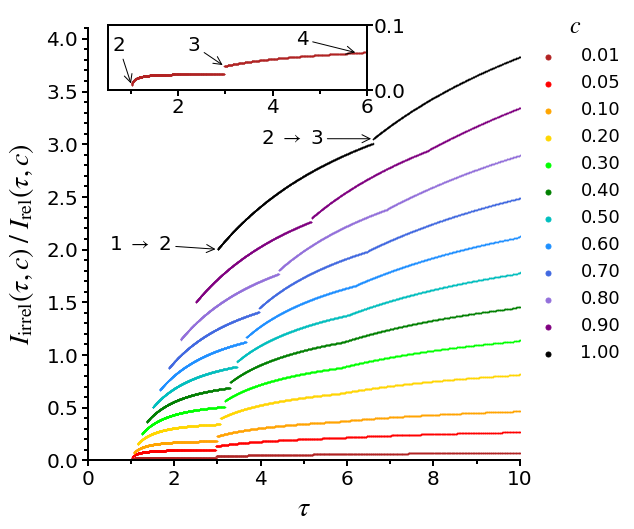}
\caption{ Ratio of irrelevant information to relevant information, evaluated for optimal memories, as a function of $\tau$ for different tilt angles. The inset shows the curve for the smallest tilt ($c=0.01$) with the corresponding phase transitions up to $K=4$ (the number of states after the transition is shown).}
\label{Irrel/Rel}
\end{figure}

The relative efficiency reduction plotted in figure \ref{Eta} has an inverse dependence on $\tau$, multiplied by the irrelevance ratio, $I_{\rm irrel}/I_{\rm rel}$: $(\eta_C-\eta)/\eta_C = (I_{\rm irrel}/I_{\rm rel})/(\tau - 1)$. In figure \ref{Irrel/Rel}, we plot the irrelevance ratio, ${I_{\rm irrel}(\tau, c) / I_{\rm rel}(\tau, c)}$ as a function of $\tau$, for different geometries (different $c$ values).

It shows that as the tilt in the geometry increases, information processing becomes a more dominant factor in reducing thermodynamic efficiency. For the geometry with the smallest tilt, at $\tau=2$, a two state memory can be found with small irrelevance ratio of approximately $I_{\rm irrel}/I_{\rm rel} = 0.02$ (see inset of figure \ref{Irrel/Rel}). As $\tau$ increases to 10, the irrelevance ratio of the corresponding optimal memory increases to approximately 0.07. In this $\tau$ regime, the inverse dependence on $\tau$ dominates the relative efficiency reduction (figure \ref{Eta}). In comparison, for the diagonally tilted geometry, at $\tau=2$, summarizing nothing at all is the best strategy, which does not allow for any work extraction, and only starting at $\tau \gtrsim 3$, do we see a phase transition to a two state memory, with irrelevance ratio of approximately 2, and a relative reduction from Carnot efficiency of nearly 100\% (figure \ref{Eta}). At $\tau=10$, the optimal memory has an irrelevance ratio of almost 4, and as a result, the engine efficiency is still reduced by over 40\%, in relative terms (figure \ref{Eta}).

The limitation posed on the quality of the observer's inference by partial observability precludes the engines from achieving Carnot efficiency. 
Other constraints could be imposed on an observer, in addition to partial observability. Another common limitation in real world situations are finite time constraints---the observer might not have enough time for (near) quasi-static operation, 
the engine might have to produce a certain amount of work in a given time frame. There is a fundamental trade-off between power and efficiency \cite{curzon1975efficiency, van2005thermodynamic, schmiedl2007efficiency, esposito2009universality, esposito2010efficiency, allahverdyan2013carnot, PhysRevLett.117.190601, PhysRevLett.120.190602, deffner2018efficiency, hong2020quantum}. Whenever time is of the essence, this trade-off should be reflected in the choice of the observer's data representation. Partial observability, as we have seen, changes the fundamental limit on achievable dissipation, and we thus expect that it also changes the 
trade-off between power and dissipation. 
Exploring physically motivated strategies for data representation in the presence of partial observability {\em and} finite time constraints, requires a stochastic thermodynamics treatment of partially observable feedback controlled systems. The insights we have presented here serve as a foundation for this extension, together with prior work on feedback controlled systems, e.g \cite{sagawa2008second, sagawa2010generalized, sagawa2012nonequilibrium, esposito2012stochastic, abreu2012thermodynamics, barato2014unifying, debiossac2020thermodynamics, cao2009thermodynamics, strasberg2019stochastic, horowitz2014second, bechhoefer2015hidden, crooks2019marginal}. Our results provide a gauge for the quality of finite time protocols: without constraints on time, intelligent observers can afford to choose minimally dissipative data representations, as an ideal strategy.

\section{Conclusion}
We introduced the first general model for partially observable information engines, via a minor modification to \LS's engine---sliding the divider into the box at an angle. This results in a partially observable \LS\ engine, since only the particle's $x$ position is observable, and needs to be used to infer which side of the box is empty.

The model allowed us to study the physical characteristics of minimally dissipative observer memories. Importantly, this opens the door to better understanding the physical basis of intelligent information processing, and of the automatization thereof. 

We used the framework of generalized information engines, where memory formation and work extraction are allowed to happen at different temperatures, because it allows for a combined treatment of heat engines and information engines. One advantage is that traditional performance measures apply, such as thermodynamic efficiency. The standard information engines typically discussed in the literature are a subclass of generalized, partially observable information engines. 
 
Minimally dissipative observer memories can be found by means of an algorithm \cite{CB, IB}. We computed those for our model class (parameterized by the tilt angle). The trade-off between a memory's thermodynamic costs versus its potential gains is governed by the temperature ratio. Using a memory with more predictive power results in more work potential; an effect proportional to the higher temperature, $T'$. The increase in the memory's thermodynamic cost is proportional to the lower temperature, $T$. Benefits of a more detailed memory thus start to outweigh costs whenever the temperature ratio $\tau = T'/T$ is large enough. For any given geometry, and for each $\tau$, the algorithm finds the minimally dissipative observer memory.

We analyzed the quality of the these optimal memories by showing how the thermodynamic gain changes as a function of the thermodynamic cost, and we discussed how the resulting curve is related to the  curve in the information plane that shows how much relevant information optimal memories keep as a function of how much total information they contain (in analogy to a rate-distortion curve). The trade-off between data compression and keeping relevant information is determined by $\tau$. We compared different geometries, which differ by their tilt angles, and thus have uncertain regions of different size. The stochastic process generating our observables (the particle's $x$ position) is less compressible (in the sense of data compression), for larger tilts, compared to smaller tilts. The solution for \LS's original engine is found trivially by the algorithm.

Memories that minimize the lower bound on dissipation (which can be achieved in the quasi-static limit) are probabilistic assignments of measurement outcomes to memory states. To gain intuition, we visualized a few of them, and compared their thermodynamic cost/benefit structure to deterministic partitions found by parametric optimization. 
We discussed a physical implementation of these probabilistic, minimally dissipative memories, and visualized the nontrivial physical ``codebook" necessary for their construction.

An information engine's thermodynamic efficiency depends on the quality with which the observer processes available information. Of the total information kept in memory, some is useful information about quantities that are relevant with respect to extracting work (here, which side of the box is empty). The remaining information is useless, or irrelevant. Memories that capture a larger relative fraction of useful, relevant information enable the engine to run at larger thermodynamic efficiency. The partial observability constraint puts a significant enough limitation on the quality of possible inferences to preclude Carnot efficiency.

\begin{acknowledgments} \noindent We thank John Bechhoefer, Josh Deutsch and Rob Shaw for extremely helpful discussions and comments. We are most grateful for funding from the Foundational Questions Institute, Grant Nos. FQXi-RFP-1820 (FQXi together with the Fetzer Franklin Fund) and FQXi-IAF19-02-S1. 
\end{acknowledgments} 

\section*{Data availability statement}
\noindent The data that support the findings of this study are available upon reasonable request from the authors.

\appendix

\renewcommand{\thesection}{\Alph{section}}
\renewcommand{\thesubsection}{\thesection.\arabic{subsection}}
\renewcommand{\thesubsubsection}{\thesubsection.\arabic{subsubsection}}
\makeatletter
\renewcommand{\p@subsection}{}
\renewcommand{\p@subsubsection}{}
\makeatother

\section{Isothermal and isentropic quasi-static transformations of a one-particle gas}
\label{thermo-basics}
In this appendix, we provide basic thermodynamic arguments to increase accessibility of the paper by saving the reader the time to find them in a textbook.
\subsection{Work during isothermal transformation, equation (\ref{W})}
Work is force times displacement. We use the convention that energy flows into the system are positive, hence work done {\em on} the gas is positive. 
Work done on an ideal gas during an isothermal, quasi-static compression performed at temperature $T$, in which the accessible volume decreases from $V_i$ to $V_f < V_i$, is given by the pressure times the volume change, $W = - \int_{V_i}^{V_f} p dV$. 
Use the ideal gas law, 
\begin{equation} 
\label{ideal-gas}
p V = N k T ,
\end{equation}
where $N$ is the number of particles, which is one here. Integrate to obtain:
\begin{equation}
W = - kT  \int_{V_i}^{V_f} {1 \over V} dV = kT \ln\left(V_i \over V_f \right) \geqslant 0.
\end{equation}
In isothermal, quasi-static {\em expansion}, the final volume is larger than the initial volume, $V_f > V_i$. Negative work is then done on the gas, equation (\ref{W}),
\begin{equation}
-W = kT \ln\left(V_f \over V_i \right)
\end{equation}
or, conversely, positive work is done by the gas on the piston, i.e., work is extracted. The sign convention differs in the literature, which can cause confusion. 

\subsection{Adiabatic condition, equation (\ref{adiabatic-eq})}
\label{TB-2}
During both isentropic transformations (compression and expansion along the $z$-axis) the work medium is isolated, no heat is exchanged with the environment. The temperature of the gas thus changes from an initial temperature, $T_i$, to a final temperature, $T_f$, as the volume of the gas changes from $V_i$ to $V_f$. Since there is no heat flow, the change in internal energy, $dE$ must equal the work done, by the first law of thermodynamics. Work done on an ideal gas is equal to the pressure times the volume change, whereby $dE = -pdV$. To express $dV$ we use the ideal gas law, equation (\ref{ideal-gas}), with $N=1$. We thus have 
\begin{eqnarray}
\label{dE-dV}
dE &=& - k {T \over V} dV.
\end{eqnarray}
The internal energy, $E$, of the work medium is given by the number of particles, $N=1$, times the average kinetic energy per particle, which is ${d \over 2} k T$, where $d$ is the number of degrees of freedom:
\begin{equation}
E =  {d\over 2} k T.
\end{equation}
For a mono-atomic gas in a container, $d$ is the dimension of the space that the particles can move in, i.e., for a 3 dimensional box, $d=3$. We can take the differential, arriving at 
\begin{eqnarray}
dE &=& {d \over 2} k dT. \label{isentr-dE}
\end{eqnarray}
With equation (\ref{dE-dV}) follows 
\begin{eqnarray}
-{1 \over V} dV &=& {d \over 2} {1 \over T} dT.
\end{eqnarray}
We integrate and obtain, for $d=3$, equation (\ref{adiabatic-eq}): 
\begin{eqnarray}
- \int_{V_i}^{V_f}{1 \over V} dV 
&=& {d \over 2} \int_{T_i}^{T_f} {1 \over T} dT \\
\Leftrightarrow  
\ln\left( {V_i \over V_f }\right)
&=& {d \over 2} \ln\left( {T_f \over T_i} \right)\\
\Leftrightarrow {V_i \over V_f } &=& \left( {T_f \over T_i} \right)^{d \over 2}. 
\end{eqnarray}

\subsection{Work during isentropic compression and expansion adds to zero.}
\label{W-cancels}
To see this, recall that $W = dE$, and use equation (\ref{isentr-dE}), integrating form an initial, $T_i$ to a final, $T_f$, temperature to arrive at
\begin{eqnarray}
W_{i\rightarrow f} &=& {d \over 2} k (T_f - T_i).
\end{eqnarray}
Thus, the work necessary to change between $T$ and $T'$ in the isentropic compression, $W_{\rm comp}$, cancels with the work done by the gas in the isentropic expansion, $W_{\rm expa}$, when the temperature changes back from $T'$ to $T$, as long as no additional degrees of freedom get unlocked at the higher temperature:
\begin{eqnarray}
W_{\rm comp}&=& {d \over 2} k (T' - T) = - W_{\rm expa}. \label{Wis0}
\end{eqnarray}

\section{Erasure protocols for \LS's engine}
\label{App-erasure}
Building on a model introduced in \cite{sagawa2009minimal}, we show here that a protocol can be constructed to assign the costs of running a memory entirely to erasure of the memory. The costs are identical to the costs for running the simple set-and-release protocol sketched in figure \ref{LSbox}. 
In  \cite{sagawa2009minimal}, and in figure \ref{LSbox}, the divider is placed in the middle of \LS's box, but the following argument can be made for arbitrary volume ratios. The calculations in this appendix use material from section \ref{LS-variants} in the main text, and it is useful to read that section before reading this appendix.
 
Let the \LS\  box have unit length and unit transverse volume. Let the container be located between $x=0$ and $x=1$, and let the divider be inserted at position $0 < \lambda \leqslant 1/2$, dividing the work medium into two separate volumes, $\lambda$ and $1-\lambda$.
The particle resides in volume $\lambda$ with probability $\lambda$, and in the other volume with probability $1-\lambda$. Knowledge of the particle's position allows for the extraction of work, equation (\ref{w-lambda}) in main text, in the amount up to (achievable in the quasi-static limit):
\begin{equation}
-W_{\rm ex}(\lambda) = - kT \left[ \lambda \ln\left({\lambda}\right) + (1- \lambda) \ln\left({1-\lambda}\right) \right] = kT h(\lambda),
\end{equation}
where $h(\lambda) \geqslant 0$ is an entropy function, as in equation (\ref{Entq}).

Let the memory be implemented by another one-particle gas in a container of unit length and unit transverse area. Let there be a parameter, $a$, controlling the relative volumes associated with the different memory states; in figure \ref{LSbox}, $a=1/2$, but in general, the memory need not be symmetric. Let $0 \leqslant x \leqslant 1$ represent the measured $x$ position of the particle, and $m \in \{0,1\}$ the two states of the memory. Note that the parametrization in this appendix is different from the one we use in the main text in section \ref{tiltedBox}, onwards. Here, the memory simply represents which side of the container is empty, i.e., all observations $0 \leqslant x \leqslant \lambda$ are mapped onto $m=0$, while  observations $\lambda < x \leqslant 1$ are mapped onto $m=1$.

First, we calculate the costs for the simple set-and-release protocol, as sketched in figure \ref{LSbox}. The ``measure and remember" step corresponds to an isothermal compression, which costs, on average, at least:
\begin{equation}
W_{\rm cost}^{\rm meas+mem}(a,\lambda) = - kT  \left[ \lambda \ln(a) + (1-\lambda) \ln(1-a) \right]. \label{cost-meas-a}
\end{equation}
This minimum can be achieved in the quasi-static limit. It depends on the choice of $a$, which should be chosen as a function of $\lambda$, in such a way that equation (\ref{cost-meas-a}) is minimized. The minimum is achieved at $a = \lambda$ (the straightforward calculation is analogous to the optimization performed in \cite{sagawa2012thermodynamics}, which was discussed after equation (\ref{w-lambda}) in the main text: the first derivative w.r.t. $a$ is set to zero, $-kT \lambda/a + kT (1-\lambda)/(1-a) \overset{!}{=} 0$, leading to $a=\lambda$; the second derivative is positive: $kT \left[\lambda/a^2 + (1-\lambda)/(1-a)^2\right] > 0$). The resulting minimal cost for this protocol is exactly equal to the maximally extractable work:
\begin{equation}
W_{\rm cost}^{\rm meas+mem}(\lambda) = kT h(\lambda) = - W_{\rm ex}(\lambda) \label{cost-meas}.
\end{equation} 
The memory is reset by simply pulling out the divider, which does not cost work, but allows for free expansion, whereby energy is dissipated.

Now, construct instead a memory manipulation protocol which contains a step of ``reset to zero", interpretable as ``erasing" the memory. (This construction is a generalization of the construction in \cite{sagawa2009minimal} to the case in which the divider in the work medium of the \LS\ box can be inserted off-center.) Start with a memory left over from the last cycle, with the divider at position $a$ (do not assume knowledge of that memory, in the sense that it is not known in which state the memory is, i.e. if the particle resides in volume $a$ or $1-a$). First, move the divider from position $a$ to position $\lambda$ isothermally and quasi-statically. With probability $\lambda$, the volume accessible to the particle is changed from $a$ to $\lambda$, and with probability $1-\lambda$, from $1-a$ to $1-\lambda$. This step serves to retrieve energy implicitly stored in memory. The average work for this step splits up into a contribution coming from the uncertainty about 
the work medium, $h(\lambda)$, 
minus
the average work done in the simple set-and-release protocol discussed above, equation (\ref{cost-meas-a}), before optimization over $a$: 
\begin{eqnarray}
&&kT \left[
\lambda \ln\left( {a \over \lambda} \right) 
+ (1-\lambda) \ln\left( {1-a \over 1-\lambda} \right)
\right] \\
&=& kT \left[ h(\lambda) + \lambda \ln(a) + (1-\lambda) \ln(1-a) \right] \\
&=& kT \left[ h(\lambda) - W_{\rm cost}^{\rm meas+mem}(a,\lambda) \right]. 
\end{eqnarray}

Second, remove the partition and compress the gas such that the particle resides in volume $a$, corresponding to the ``zero" reset state. This procedure can be interpreted as erasing the memory, cf. \cite{sagawa2009minimal}. This isothermal compression costs at least $-kT \ln(a)$. Overall, the cost of erasing the memory, by changing it from its former state to the ``zero" reset state, is then the sum:
\begin{eqnarray}
W_{\rm cost}^{\rm erase}(a, \lambda) &=& kT \left[ h(\lambda)\! -\! (1\!-\!\lambda) \ln\left( a \over 1-a \right) \right].
\end{eqnarray} 
Finally, a last step is needed to set the memory to the new value. With probability $\lambda$, the ``zero" state correctly corresponds to the new measurement outcome and does not need to be changed. With probability $1-\lambda$, it needs to be transformed into the other state. The costs encountered for this transformation come from changing the volume the particle resides in from $a$ to $1-a$. Thus, this last step, which can be interpreted as setting the memory, costs on average at least 
\begin{eqnarray}
W_{\rm cost}^{\rm set}(a, \lambda) &=& kT (1-\lambda) \ln\left( {a\over 1-a} \right).
\end{eqnarray}  
Combining all costs for this protocol, we see that its minimum (quasi-static) costs are exactly the same as those of the simpler set-and-release protocol, again, cancelling out all possible gains one could derive 
over a cycle:
\begin{equation}
W_{\rm cost}^{\rm erase+set}(\lambda) = kT h(\lambda) = W_{\rm cost}^{\rm meas+mem}(\lambda) = - W_{\rm ex}(\lambda).
\end{equation}
However, the way in which $W_{\rm cost}^{\rm erase+set}(\lambda)$ has been split up is still dependent on the parameter $a$, which can be chosen at will, without changing the overall costs of the full protocol. One choice of $a$ would attribute all costs to the ``erasure" by ensuring $W_{\rm cost}^{\rm set}(a, \lambda) = 0$; this choice is $a=1/2$, corresponding to a symmetric memory. Another choice would make the erasure costs zero, $W_{\rm cost}^{\rm erase}(a, \lambda) =0$, and attribute all costs to the setting of the memory. This choice is realized by 
\begin{equation}
a = { e^{h(\lambda)/(1-\lambda)} \over 1+ e^{h(\lambda)/(1-\lambda)}}.
\end{equation}

\section{Calculation of equation (\ref{HUgX})}
\label{AppA}
The probability density of the particle's $x$ position, $p(x)$, is constant inside the container, and the length of the container is set to unity, so to calculate $- H[U|X]$, we need to evaluate the integral 
\begin{eqnarray}
\int_{-{c/2}}^{{c/2}} \left[ \left({1\over 2} \!+\! {x \over c} \right) \ln\left({1\over 2} \!+\! {x \over c} \right) \!+\! \left({1\over 2} \!-\!{x \over c} \right) \ln\left({1\over 2} \!-\! {x \over c} \right)\right] dx.  \notag 
\end{eqnarray}
In the first term, substitute $x' = {1\over 2} + {x \over c}$, which gives a factor $dx = c dx'$, and for the second term substitute $x'' = {1\over 2} -{x \over c}$, with $dx = -c dx''$. Note that the limits on the integrals are then just reversed, and because of the negative sign on the second term, both terms are the same. Let us choose $y$ for the new integration variable: 
\begin{eqnarray}
- H[U|X] 
&=& c \int_{0}^{1} x' \ln(x') dx' - c \int_{1}^{0} x''  \ln(x'') dx'' \notag \\
&=& {2c} \int_{0}^{1} y \ln(y) dy.
\end{eqnarray}
We solve the indefinite integral by parts,
\begin{eqnarray}
\int \; y \ln(y)  \; dy &=& {1 \over 2} y^2 \ln(y) - {1 \over 2} \int y dy\\
&=& {1 \over 2} y^2 \left( \ln(y) - {1 \over 2} \right), 
\end{eqnarray}
and evaluate the limits to arrive at equation (\ref{HUgX}):
\begin{equation}
- H[U|X] = - {c \over 2}.
\end{equation}

\section{Boxes with $\pi/4 \geqslant \theta \geqslant 0$}
\label{flat-tilt-boxes}
\begin{figure}[h]
\centering 
\includegraphics[width=0.75\linewidth]{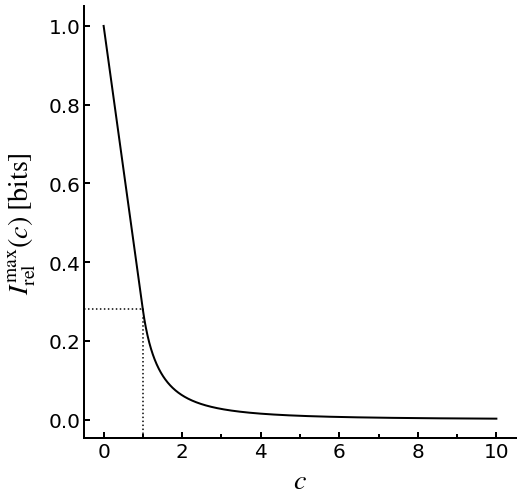}
\caption{Maximum relevant information $I_{\rm rel}^{\rm max}(c)$. Dotted lines mark $c=1$ with $I_{\rm rel}^{\rm max}(c=1) \approx 0.28$ bits.}
\label{fig:max_irel}
\end{figure}

If we do the same calculation for $c\geqslant 1$, the integration limits change, and we have to compute $- H[U|X]=$
\begin{eqnarray}
\int_{-{1/2}}^{{1/2}} \left[ \left({1\over 2} \!+\! {x \over c} \right) \ln\left({1\over 2} \!+\! {x \over c} \right) \!+\! \left({1\over 2} \!-\!{x \over c} \right) \ln\left({1\over 2} \!-\! {x \over c} \right)\right] dx  \notag 
\end{eqnarray}
After the coordinate transformations we have 
\begin{eqnarray}
- H[U|X] 
&=& {2c} \int_{{1\over 2} \left( 1-{1\over c}\right) }^{{1\over 2} \left( 1+{1\over c}\right) } y \ln(y) dy,\\
&=& c \; y^2 \left[ \ln(y) - {1 \over 2} \right)\Bigg|_{{1\over 2} \left( 1-{1\over c}\right)}^{{1\over 2} \left( 1+{1\over c}\right] } 
\end{eqnarray}
and, altogether, we arrive at 
\begin{eqnarray}
- H[U|X] 
&=&-\ln(2) -{1\over 2} + {c\over 4} \ln{\left({c+1 \over c-1}\right)} \notag  \\
&&+ {1\over 4c} \ln{\left({c+1 \over c-1}\right)}\!+\! {1\over 2} \ln\left(1\!-\!{1\over c^2}\right)\\
&\stackrel{c\rightarrow \infty}{\longrightarrow}& - \ln(2), 
\end{eqnarray}
because $c \ln{\left({c+1 \over c-1}\right)} \stackrel{c\rightarrow \infty}{\longrightarrow} 2$, as one can see by using L'Hopital's rule. As expected, the maximum relevant information goes to zero as we tilt the divider closer to parallel with the $x$-axis. We plot $I_{\rm rel}^{\rm max}(c)$ as a function of $c$ in figure \ref{fig:max_irel}.

\begin{figure}[h]
\centering 
\includegraphics[width=0.75\linewidth]{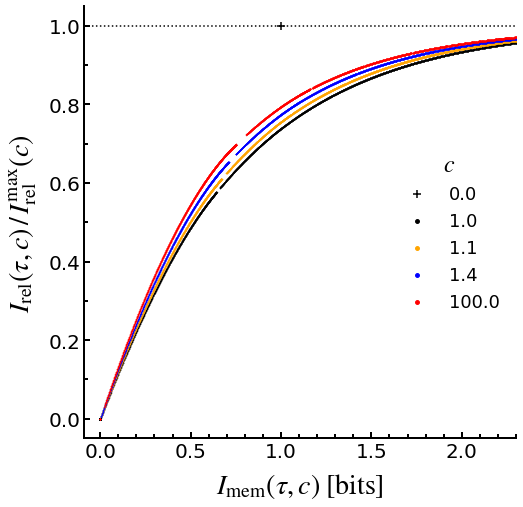}
\caption{ Information plane curves of optimal data representations for selected $c \geqslant 1$.}
\label{fig:largec}
\end{figure}

We choose to discuss only geometries up to $c=1$ ($\theta=\pi/4$) in the main text, because for the larger tilts nothing qualitatively different happens. For completeness, we show in figure \ref{fig:largec}  $I_{\rm rel}(\tau,c)/I_{\rm rel}^{\rm max}(c)$ as a function of $I_{\rm mem}(\tau,c)$ at the optimal data representations for geometries with selected $c \geqslant 1$. 

\section{Calculation of $I_{\rm rel}^{\rm cer} / I_{\rm rel}^{\rm max} = (1 - c) / [1 - c/ \ln(4)]$, displayed in figure \ref{rel-cont-cert}} 
\label{App:rel_certain}
Only the uncertain region contributes to the conditional entropy. The certain region's contribution to the relevant information is:  
\begin{eqnarray}
 \!\!\! \!\!\! \!\!\! I_{1-c}[X,U] &=& 2 \int_{{c\over 2}}^{{1\over 2}} dx \sum_{u\in \{-1,1\}} p(x,u) \ln{ \left[ {p(u|x) \over p(u)} \right] }\\
&=& 2 \int_{{c\over 2}}^{{1\over 2}} dx  \sum_{u\in \{-1,1\}} \cancelto{0}{p(u|x) \ln\left[ p(u|x) \right]} \notag \\
&& + 2 \, \ln(2) \int_{{c\over 2}}^{{1\over 2}} dx  \cancelto{1}{\sum_{u\in \{-1,1\}} p(u|x)}  \notag\\
&=&  \left( 1 - c \right)  \ln(2) 
\end{eqnarray} 

The relative contribution, plotted in figure \ref{rel-cont-cert}, is 
\begin{eqnarray}
{ I_{1-c}[X,U] \over  I[X,U] } 
&=&   {\ln(2)  \left( {1} - {c} \right) \over \ln(2) - {c\over 2} } =  {{1} - {c} \over 1 - {c\over \ln(4)}}.
\end{eqnarray} 

\section{Comparison to boxes with fixed angle and varying size of the certain regions}
\label{comp-varyL-boxes}
Note that we do not lose generality by assuming unit container length in $x$-direction. Were we to keep the length as a parameter, $L$, then we would simply have to replace $c$ in all the calculations above by $c/L$, the relative length of the uncertain region. 

Two boxes with different tilt angles, and thus different $c_1$ and $c_2$, and different lengths $L_1$ and $L_2$, respectively, behave the same whenever the relative length of the uncertain region is the same, i.e. $c_1/L_1 = c_2/L_2 = c/L$. We checked this numerically: solutions to the optimization problem equation (\ref{IB}) are the same. 
\begin{figure}[h]
\centering 
\includegraphics[width=0.8\linewidth]{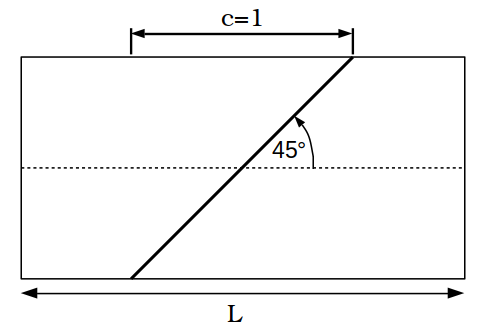}
\caption{Equivalent parametrization of the partially observable \LS\ engine. The angle is fixed to $\theta = \pi/4$ and the length of the box along the $x$-axis, $L$, is varied.}
\label{LargerBox}
\end{figure} 

We could thus have used an alternative, equivalent parametrization: fix the angle to, for example, $\theta = \pi/4$, i.e. $c=1$, and let the container length along the $x$-axis, $L$, vary. This is sketched in figure \ref{LargerBox}. What matters is the relative length of the certain regions, $L-c$, compared to the uncertain region, $c$: $(L-c)/c = L/c-1$.

\section{Deterministic memories, computed from parametric minimization of dissipation}
\label{Det-mindiss}
\begin{figure}[h]
\centering 
\includegraphics[width=0.8\linewidth]{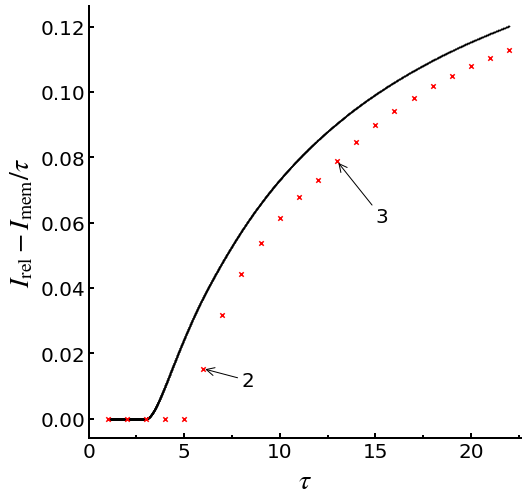}
\caption{ Total work output in units of $kT'$, as a function of $\tau$, comparing deterministic memories (red) to optimal probabilistic memories (black). Numbers point to the first solutions with two and three states.}
\label{objfkt}
\end{figure}

We can allow for deterministic partitions of any shape (including asymmetric solutions), where the number of memory states is a free parameter. For the box with diagonal tilt, $c=1$, we maximize work output, equation (\ref{Woutengine}), over these deterministic memories, to compare their performance to the minimally dissipative {\it probabilisitic} memories we found by solving optimization equation (\ref{IB}). Figure \ref{objfkt}, shows the total work output, in units of $kT'$, as a function of the temperature ratio $\tau$. 

Compared to the optimal, minimally dissipative memories, deterministic memories achieve comparable work output at higher values of $\tau$. To obtain nonzero output, the temperature ratio has to be larger by roughly 1.75.
\begin{figure}[h]
\centering 
\includegraphics[width=0.8\linewidth]{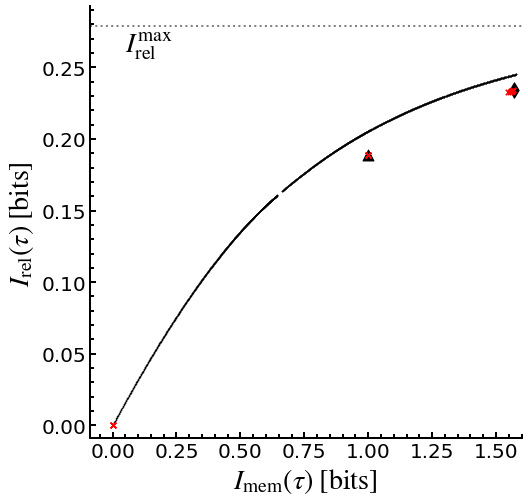}
\caption{ Comparing deterministic memories (red) to optimal probabilistic memories (black) in the information plane, and to deterministic memories found as $\tau \rightarrow \infty$, when partitions are constrained to two (black triangle) and three (diamond) number of states.}
\label{I-plane-comp}
\end{figure} 

\begin{figure}[h]
\centering 
\includegraphics[width=0.8\linewidth]{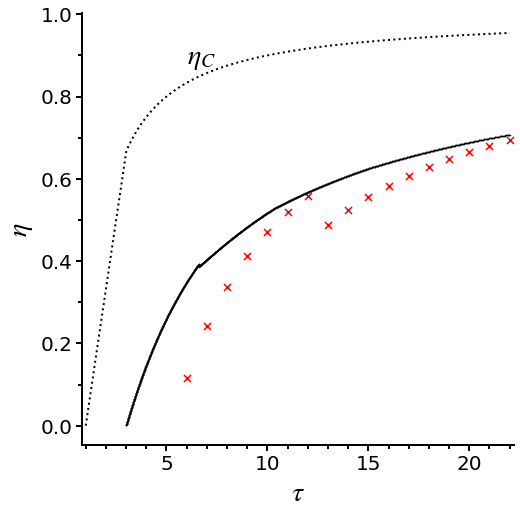}
\caption{ Efficiency achieved by deterministic memories (red) compared to optimal probabilistic memories (black), as a function of $\tau$. Carnot efficiency, $\eta_C$, is traced out as a dotted line.}
\label{eta-comp}
\end{figure}

The data are re-displayed in the information plane in figure \ref{I-plane-comp}, to reveal that the deterministic memories are close to those found for $\tau \rightarrow \infty$, when partitions are constrained to a fixed number of states. The closer they are to these solutions, the closer they come to the efficiency that optimal memories afford, as displayed in figure \ref{eta-comp} \footnote{In this exploration, we broke off the numerical optimization at the point where we observed a transition to four memory states, because the trends are clearly visible at that point. We thus show only solutions up to three memory states. We calculated parametrically optimized deterministic partitions at $\tau$ intervals of 1, which is coarser than the spacing used to compute optimal memories.}. 

\section{Suboptimal four- and five-state memories}
\label{AppB}
To show that the deterministic four- and five-state memories discussed in the main text, section \ref{45s} are optimal, we discuss the suboptimal cases here.

The simplest four-state memory one can think of maps the certain regions to two distinct states and splits the uncertain region into two more states (table \ref{table:4s}).
\begin{table}[h]
\begin{tabular}{l|c|l} 
$x$ & $m$ & $q_4^{d=c}(m)$ \\ [0.5ex]
\hline
$-1/2 \leqslant x \leqslant -c/2$ & $-2$ & $0$ \\ [0.5ex]
$-c/2 < x \leqslant 0$ & $-1$ & $1/4$ \\ [0.5ex]
$0 < x \leqslant c/2$ & $1$ & $1/4$ \\ [0.5ex]
$c/2 < x \leqslant 1/2$ & $2$ & $0$
\end{tabular}
\caption{Memory assignments and errors for a deterministic four-state memory with $d=c$.}
\label{table:4s}
\end{table}

This coarse graining has two certain states, $q_4^{d=c}(m=\pm 2) = 0$, and two states with the same fixed error probability, $q_4^{d=c}(m=\pm 1) = 1/4$. As in the main text, the error is found by dividing the area of the triangle on the other side of the divider, which is $c/8$ by the total area of the $\pm1$ state, which is $c/2$.

For this memory the relevant information is given by
\begin{eqnarray}
I_{\rm rel}^{(4)} (d=c) &=& (1-c)\ln(2) + c\,[\ln(2)-h(1/4)] \notag \\
&=& \ln(2) - c \, h(1/4).
\end{eqnarray}

The second type of suboptimal symmetrical four-state memory has the same state assignments as the optimal deterministic four-state memory (table \ref{table:4sd}) but it has larger $m=\pm 1$ states that include parts of the certain region ($d > c$). Therefore the errors are different and we find $q_4^{d>c}(m=\pm 2) = 0$ and $q_4^{d>c}(m=\pm 1) = {c \over 4d}$. For the three-state memory it was clear that choosing $\ell > c$ decreases the amount of relevant information captured since it simply increases the size of the $m=0$ state that cannot be used for work extraction. For four-state memories it is not immediately obvious that $d > c$ is a similarly poor choice. While the size of the certain states decreases from $1-c$ to $1-d$, the error for the $m=\pm 1$ states also decreases. The relevant information for this assignment is given by:
\begin{eqnarray} \label{4slarge}
I_{\rm rel}^{(4)}(d>c) &=& (1-d)\ln(2) + d\left[\ln(2) - h\!\left({c \over 4d}\right)\right] \notag \\
&=&\ln(2) - d \, h\!\left({c \over 4d}\right).
\end{eqnarray}
For five-state memories there are two parametrizations with one parameter equal to $c$ ($d = c$ or $\ell = c$). For the first ($d=c$, table \ref{table:5l}) the size of the waste basket state $m=0$ is given by $\ell < c$:
\begin{table}[h]
\begin{tabular}{ l|c|l } 
$x$ & $m$ & $q_5^{\ell<c}(m)$ \\ [0.5ex]
\hline
$-1/2 \leqslant x \leqslant -c/2$ & $-2$ & $0$ \\ [0.5ex]
$-c/2 < x \leqslant -\ell/2$ & $-1$ & $\tilde{\epsilon}(\ell, c)$ \\ [0.5ex]
$-\ell/2 < x \leqslant \ell/2$ & $0$ & $1/2$ \\ [0.5ex]
$\ell/2 < x \leqslant c/2$ & $1$ & $\tilde{\epsilon}(\ell, c)$ \\ [0.5ex]
$c/2 < x \leqslant 1/2$ & $2$ & $0$
\end{tabular}
\caption{Memory assignments and errors for a five-state memory with $\ell < c$ and $d=c$.}
\label{table:5l}
\end{table}
with $\tilde{\epsilon}(\ell, c) = \left({c-\ell \over 4c}\right)$. For this assignment the relevant information is given by:
\begin{eqnarray}
I_{\rm rel}^{(5)}(\ell\!<\!c, d=c) &=& (1\!-\!c) \ln(2) \!+\! (c\!-\!\ell)\left[\ln(2)\!-\!h(\tilde{\epsilon})\right] \notag \\
&=& (1\!-\!\ell)\ln(2) \!-\! (c\!-\!\ell)\,h(\tilde{\epsilon}(\ell, c)).
\end{eqnarray}
For the second option ($\ell=c$) the states $m=\pm 1$ have $q_5^{\ell=c}(m=\pm 1)=0$ since they are completely within the certain region for $d > \ell = c$. Clearly this coarse graining then captures the same amount of relevant information as a three-state memory with $\ell = c$. The two certain states are further subdivided into 2 states each, but this only increases the cost of running the memory while not capturing any additional relevant information:
\begin{eqnarray}
I_{\rm rel}^{(5)}(l=c,d\!>\!c) &=& (1-\ell)\ln(2) + (\ell-c)\ln(2) \notag \\
&=& (1-c)\ln(2) = I_{\rm rel}^{(3)}(\ell=c).
\end{eqnarray}
More interesting are the five-state memories with both parameters unequal to $c$. Choosing $\ell > c$ only leads to subdividing the certain regions into two states each and is clearly an even poorer choice than $\ell=c$. For smaller waste baskets, $\ell < c$, the option not discussed in the main text (section \ref{45s}) cuts into the certain regions, $d > c$. This parametrization is presented in table \ref{table:5ld_large}.
\begin{table}[h]
\begin{tabular}{ l|c|l } 
$x$ & $m$ & $q_5^{c < d}(m)$ \\ [0.5ex]
\hline
$-1/2 \leqslant x \leqslant -d/2$ & $-2$ & $0$ \\ [0.5ex]
$-d/2 < x \leqslant -\ell/2$ & $-1$ & $\bar{\epsilon}(\ell, c, d)$ \\ [0.5ex]
$-\ell/2 < x \leqslant \ell/2$ & $0$ & $1/2$ \\ [0.5ex]
$ \ell/2 < x \leqslant d/2$ & $1$ & $\bar{\epsilon}(\ell, c, d)$ \\ [0.5ex]
$ d/2 < x \leqslant 1/2$ & $2$ & $0$
\end{tabular}
\caption{Memory assignments and errors for a five-state memory with $\ell < c$ and $d > c$.}
\label{table:5ld_large}
\end{table}
The error for the states $m=\pm 1$ is given by
\begin{equation}
\bar{\epsilon}(\ell,c,d) = {(c-\ell)^2 \over 4c(d-\ell)}.
\end{equation}
The relevant information captured by this memory is
\begin{eqnarray} \label{5slarged}
I_{\rm rel}^{(5)}(\ell\!<\!c, c\!<\!d) &=& \! (1\!-\!d)\ln(2)\! +\! (d\!-\!\ell)\left[\ln(2)\!-\!h(\bar{\epsilon})\right] \notag \\
&=& \! (1\!-\! \ell)\ln(2) \! - \! (d \!-\! \ell)\,h(\bar{\epsilon}(\ell,c,d)).
\end{eqnarray}
For $\ell \rightarrow 0$ we recover equation (\ref{4slarge}). 

\begin{figure}[h]
\centering 
\includegraphics[width=0.8\linewidth]{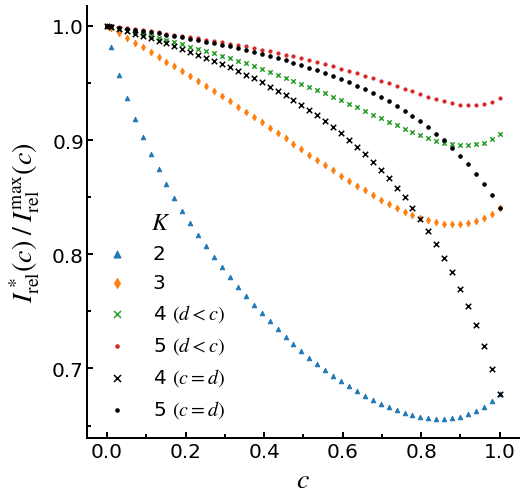}
\caption{Fraction of the maximum relevant information captured in memory, for optimal compared to suboptimal parametrizations, as a function of $c$.}
\label{optimalIfull}
\end{figure}
Just as for three-state memories (section \ref{3s}), splitting the certain region into more than two memory states ($d > c$) is the worst choice for symmetric four- and five-state memories. Decreasing the error of the $m=\pm1$ states can never compensate having smaller certain states. Consequently for parametrizations with $d > c$ the best partitions will have their certain states as larges as possible, i.e. $d$ as close to $c$ as possible.
In figure \ref{optimalIfull} we see that it is also never beneficial to choose $d = c$ in four- and five-state memories since memories with the same number of states but $d < c$ are able to retain a larger proportion of the maximum relevant information available for a given geometry.

\section{Compressibility (adopted from \cite{StillCru07})}
\label{compress}
A data generating process (a source producing $X$) is said to be {\em fully} compressible if the information curve $(I_{\rm mem}(\tau), I_{\rm rel}(\tau))$ traced out for all optimal data representations (where optimality is understood in terms of minimizing wasted energy, i.e. the solutions to equation (\ref{IB})) reaches the maximum, at which $I_{\rm rel}(\tau) = I_{\rm rel}^{\rm max}$, for a memory that achieves a compression of the input data, i.e. for which $I_{\rm mem}(\tau) < H[X]$. 

Any source with a curve above the diagonal spanned by (0,0) and $(H[X], I_{\rm rel}^{\rm max})$ is compressible. The smaller $I_{\rm mem}(\tau)$ is at the point 
where $I_{\rm rel}(\tau)$ asymptotically reaches $I_{\rm rel}^{\rm max}$, i.e. the larger the curvature, the more compressible is the source. 
Applied to figure \ref{IBcurves}, geometries for which smaller memories contain a larger fraction of relevant information are thus said to be more compressible.

\section{Information Bottleneck Algorithm Implementation}
\label{AppC}
Numerical calculations are performed using the Information Bottleneck algorithm together with a ``deterministic annealing" procedure adapted from \cite{rose1998deterministic}. Starting with $\tau = 1$, for each value of $\tau$, the self-consistent equations (equations (\ref{pmx})-(\ref{pum})) are iterated until convergence (algorithm \ref{EMstepcode}). Then, $\tau$ is slowly increased. The solution at the previous value for $\tau$ is handed to the algorithm as an initial condition for the iterations at the next value of $\tau$. 
During this procedure, the size of the memory grows. This is implemented by doubling the number of memory states and perturbing them slightly. Then the effective number of states is found by inspecting the distributions $p(u|m)$ after the iterative equations converge. If they are the same for two different values of $m$, then the corresponding states are merge together. 
The algorithm finishes when a maximum temperature ratio $\tau_{\rm max}$ is reached (or when the solutions for a fixed number of states $K_{\rm max}$ are deterministic), a practical choice to cut down computational time. In principle we could let it run until there is one memory state for every $x$-value, the point at which all data is recalled, and the memory is no longer a summary. Since we are not particularly interested in this asymptotic regime, we end early.
Numerical calculations are performed using algorithm \ref{IBcode}. Algorithm \ref{EMstepcode} describes a subroutine used in algorithm \ref{IBcode}. 

\onecolumngrid

\begin{algorithm}[H]
    \caption{Iterate self-consistent Information Bottleneck equations}
    \label{EMstepcode}
    {\fontsize{10}{30} \begin{algorithmic}
    \State \textbf{Require:} $p(x,u)$; joint probability distribution of observables $x$ and relevant variables $u$
    \State \textbf{Require:} $p(m|x)$; probabilistic assignments of observables $x$ to memory states $m$ 
    \State \textbf{Require:} $p(m)$; probability for each memory state $m$ 
    \State \textbf{Require:} $p(u|m)$; probability of relevant variable having realization $u$, given memory state $m$
    \State \textbf{Require:} $\tau={T' / T}$; temperature ratio between work extraction step ($T'$) and memory making step ($T$)
        \While{$p(u|m)$ not converged}
            \vspace*{3pt}
            \State Update $p(m|x) \leftarrow p(m) \exp(-\tau \, {\cal D}(p(u|m) \| p(m|x)))$
            \vspace*{3pt}
        	\State Normalize $p(m|x) \leftarrow \frac{p(m|x)}{\sum_m p(m|x)}$
        	\vspace*{3pt}
        	\State Update $p(m) \leftarrow \sum_x p(m|x) p(x)$
        	\vspace*{3pt}
        	\State Update $p(u|m) \leftarrow \frac{\sum_x p(x,u) p(m|x)}{p(m)}$
        	\EndWhile
    \vspace*{3pt}
    \State Compute: $I[M,X] \leftarrow \sum_{x,m} p(m|x)p(x) \log\left(\frac{p(m|x)}{p(x)}\right)$, $I[U,M] \leftarrow \sum_{u,m} p(u|m)p(m) \log\left(\frac{p(u|m)}{p(u)}\right)$
    \State \textbf{return} $\tau$, $I[M,X]$, $I[U,M]$, $p(m|x)$, $p(u|m)$, $p(m)$.
    \end{algorithmic}}
\end{algorithm}

\twocolumngrid

\onecolumngrid

\begin{algorithm}[H]
\caption{Information Bottleneck with annealing}
\label{IBcode}
\begin{algorithmic}
\State \textbf{Require} $p(x,u)$; joint probability distribution of observables $x$ and relevant variables $u$ 
\State \textbf{Require} $\tau_{\rm init}$, $\tau_{\rm max}$; initial and maximum value for $\tau$
\State \textbf{Require} $\alpha$; annealing rate for increasing $\tau$ after each iteration 
\State \textbf{Require} $perturb$; maximum perturbation to be added when splitting memory states
\State \textbf{Require} $merge$; minimum element-wise difference for memory states not to be merged
\State \textbf{Require} $K_{\rm max}$; maximum number of memory states (for subroutine finding deterministic assignments) 
\State Initialize: $K = 1$, $\tau = \tau_{\rm init}$, $p(m|x) = 1/K, \forall x$
\State Compute: $p(x) = \sum_u p(x,u)$, $p(u|x) = \frac{p(x,u)}{p(x)}$, $p(m) = \sum_x p(m|x) p(x)$, $p(u|m) = \frac{\sum_x p(x,u) p(m|x)}{p(m)} $
\While{$K < K_{\rm max}$ \textbf{and} $\tau < \tau_{\rm max}$}
\State $\tau \leftarrow \alpha \cdot \tau$
\State Split $K \rightarrow 2K$ and add random perturbation ($< perturb$) to each memory state
\State \textbf{Iterate self-consistent Information Bottleneck equations} until convergence 							for $K=2K$
\State Merge memory states $2K \rightarrow K'$ ($K' \in [K, K+1, K+2,..., 2K]$)
\State \textbf{Iterate self-consistent Information Bottleneck equations} until convergence 							for $K=K'$
\State Store $\tau$, 
$I_{\rm mem} \leftarrow I[M,X]$, $I_{\rm rel} \leftarrow I[U,M]$
\EndWhile
\State \textbf{Subroutine:} Find deterministic memory assignments (assignments are either 0 or 1) for $K=K_{\rm max}$
\While{$p(m|x)$ not deterministic \textbf{and} $p(m)$ not converged \textbf{and} $\tau < 								\tau_{\rm max}$}
\State $\tau = \alpha \cdot \tau$
\State Iterate self-consistent equations until convergence
\EndWhile
\State \textbf{return} $\tau$, $I_{\rm mem}$, $I_{\rm rel}$
\end{algorithmic}
\label{alg1}
\end{algorithm}

\twocolumngrid

\bibliographystyle{unsrt}
\bibliography{POSE_citations}

\begin{thebibliography}{10}

\bibitem{maxwell-demon}
J.~C. Maxwell.
\newblock {Letter to P. G. Tait, 11 December 1867}.
\newblock In C.~G. Knott, editor, {\em Life and Scientific Work of Peter
  Guthrie Tait}, page p. 213. Cambridge University Press, London, 1911.

\bibitem{maxwell1871theory}
J.~C. Maxwell.
\newblock {\em Theory of Heat}.
\newblock Text-books of science. Longmans, 1871.

\bibitem{szilard}
L.~Szilard.
\newblock On the decrease of entropy in a thermodynamic system by the
  intervention of intelligent beings.
\newblock {\em Z. Phys.}, 53:840--856, 1929.

\bibitem{Landauer1961}
R.~Landauer.
\newblock Irreversibility and heat generation in the computing process.
\newblock {\em IBM J. Res. Develop.}, 5(3):183--191, 1961.

\bibitem{demonbook90}
A.~Rex and H.~S. Leff, editors.
\newblock {\em {Maxwell's Demon: Entropy, Information, Computing}}.
\newblock A. Hilger (Bristol, England), 1990.

\bibitem{bennett2003notes}
C.H. Bennett.
\newblock {Notes on Landauer's principle, reversible computation, and Maxwell's
  Demon}.
\newblock {\em Studies In History and Philosophy of Science Part B: Studies In
  History and Philosophy of Modern Physics}, 34(3):501--510, 2003.

\bibitem{Leff2003}
H.~S. Leff and A.~F. Rex, editors.
\newblock {\em {Maxwell's Demon 2: E}ntropy, Classical and Quantum Information,
  Computing}.
\newblock IOP Publishing, Bristol and Philadelphia, 2003.

\bibitem{toyabe2010experimental}
S.~Toyabe, T.~Sagawa, M.~Ueda, E.~Muneyuki, and M.~Sano.
\newblock {Experimental demonstration of information-to-energy conversion and
  validation of the generalized Jarzynski equality}.
\newblock {\em Nat. Phys.}, 6(12):988--992, 2010.

\bibitem{berut2012experimental}
A.~B{\'e}rut, A.~Arakelyan, A.~Petrosyan, S.~Ciliberto, R.~Dillenschneider, and
  E.~Lutz.
\newblock {Experimental verification of Landauer’s principle linking
  information and thermodynamics}.
\newblock {\em Nature}, 483(7388):187--189, 2012.

\bibitem{koski2014experimental}
J.~V. Koski, V.~F. Maisi, T.~Sagawa, and J.~P. Pekola.
\newblock {Experimental Observation of the Role of Mutual Information in the
  Nonequilibrium Dynamics of a Maxwell Demon}.
\newblock {\em Phys. Rev. Lett.}, 113:030601, Jul 2014.

\bibitem{koski2014SEszilard}
J.~V. Koski, V.~F. Maisi, J.~P. Pekola, and D.~V. Averin.
\newblock {Experimental realization of a Szilard engine with a single
  electron}.
\newblock {\em Proc. Natl. Acad. Sci. U.S.A.}, 111(38):13786--13789, 2014.

\bibitem{exp-landauer2014}
Y.~Jun, M.~Gavrilov, and J.~Bechhoefer.
\newblock {High-Precision Test of Landauer's Principle in a Feedback Trap}.
\newblock {\em Phys. Rev. Lett.}, 113:190601, Nov 2014.

\bibitem{koski2015chip}
J.~V. Koski, A.~Kutvonen, I.~M. Khaymovich, T.~Ala-Nissila, and J.~P. Pekola.
\newblock {On-chip Maxwell’s demon as an information-powered refrigerator}.
\newblock {\em Phys. Rev. Lett.}, 115(26):260602, 2015.

\bibitem{martinez2016brownian}
I.~A. Mart{\'\i}nez, E.~Rold{\'a}n, L.~Dinis, D.~Petrov, J.~M.~R. Parrondo, and
  R.~A. Rica.
\newblock {Brownian Carnot engine}.
\newblock {\em Nat. Phys.}, 12(1):67--70, 2016.

\bibitem{hong2016experimental}
J.~Hong, B.~Lambson, S.~Dhuey, and J.~Bokor.
\newblock {Experimental test of Landauer's principle in single-bit operations
  on nanomagnetic memory bits}.
\newblock {\em Science advances}, 2(3):e1501492, 2016.

\bibitem{camati2016experimental}
P.~A. Camati, J.~P.~S. Peterson, T.~B. Batalhao, K.~Micadei, A.~M. Souza, R.~S.
  Sarthour, I.~S. Oliveira, and R.~M. Serra.
\newblock {Experimental rectification of entropy production by Maxwell’s
  demon in a quantum system}.
\newblock {\em Phys. Rev. Lett.}, 117(24):240502, 2016.

\bibitem{gavrilov2016erasure}
M.~Gavrilov and J.~Bechhoefer.
\newblock Erasure without work in an asymmetric double-well potential.
\newblock {\em Phys. Rev. Lett.}, 117:200601, Nov 2016.

\bibitem{gavrilov2017direct}
M.~Gavrilov, R.~Ch{\'e}trite, and J.~Bechhoefer.
\newblock {Direct measurement of weakly nonequilibrium system entropy is
  consistent with Gibbs{\textendash}Shannon form}.
\newblock {\em Proc. Natl. Acad. Sci. U.S.A}, 114(42):11097--11102, 2017.

\bibitem{chida2017power}
K.~Chida, S.~Desai, K.~Nishiguchi, and A.~Fujiwara.
\newblock {Power generator driven by Maxwell’s demon}.
\newblock {\em Nat. Commun.}, 8(1):1--7, 2017.

\bibitem{cottet2017observing}
N.~Cottet, S.~Jezouin, L.~Bretheau, P.~Campagne-Ibarcq, Q.~Ficheux, J.~Anders,
  A.~Auff{\`e}ves, R.~Azouit, P.~Rouchon, and B.~Huard.
\newblock {Observing a quantum Maxwell demon at work}.
\newblock {\em Proc. Natl. Acad. Sci. U.S.A.}, 114(29):7561--7564, 2017.

\bibitem{ciampini2017experimental}
M.~A. Ciampini, L.~Mancino, A.~Orieux, C.~Vigliar, P.~Mataloni, M.~Paternostro,
  and M.~Barbieri.
\newblock Experimental extractable work-based multipartite separability
  criteria.
\newblock {\em npj Quantum Information}, 3(1):1--6, 2017.

\bibitem{kumar2018nanoscale}
A.~Kumar and J.~Bechhoefer.
\newblock Nanoscale virtual potentials using optical tweezers.
\newblock {\em Appl. Phys. Lett.}, 113(18):183702, 2018.

\bibitem{paneru2018lossless}
G.~Paneru, D.~Y. Lee, T.~Tlusty, and H.~K. Pak.
\newblock Lossless brownian information engine.
\newblock {\em Phys. Rev. Lett.}, 120(2):020601, 2018.

\bibitem{admon2018experimental}
T.~Admon, S.~Rahav, and Y.~Roichman.
\newblock Experimental realization of an information machine with tunable
  temporal correlations.
\newblock {\em Phys. Rev. Lett.}, 121(18):180601, 2018.

\bibitem{ribezzi2019large}
M.~Ribezzi-Crivellari and F.~Ritort.
\newblock {Large work extraction and the Landauer limit in a continuous Maxwell
  demon}.
\newblock {\em Nat. Phys.}, 15(7):660--664, 2019.

\bibitem{peterson2020implementation}
J.~P.~S. Peterson, R.~S. Sarthour, and R.~Laflamme.
\newblock Implementation of a quantum engine fuelled by information.
\newblock {\em arXiv:2006.10136}, 2020.

\bibitem{paneru2020efficiency}
G.~Paneru, S.~Dutta, T.~Sagawa, T.~Tlusty, and H.~K. Pak.
\newblock Efficiency fluctuations and noise induced refrigerator-to-heater
  transition in information engines.
\newblock {\em Nat. Commun.}, 11(1):1--8, 2020.

\bibitem{paneru2020colloidal}
G.~Paneru and H.~K. Pak.
\newblock Colloidal engines for innovative tests of information thermodynamics.
\newblock {\em Advances in Physics: X}, 5(1):1823880, 2020.

\bibitem{saha2020maximizing}
T.~K. Saha, J.~N.~E. Lucero, J.~Ehrich, D.~A. Sivak, and J.~Bechhoefer.
\newblock Maximizing power and velocity of an information engine.
\newblock {\em Proc. Natl. Acad. Sci. U.S.A.}, 118(20), 2021.

\bibitem{dago2021information}
S.~Dago, J.~Pereda, N.~Barros, S.~Ciliberto, and L.~Bellon.
\newblock Information and thermodynamics: Fast and precise approach to
  landauer’s bound in an underdamped micromechanical oscillator.
\newblock {\em Phys. Rev. Lett.}, 126(17):170601, 2021.

\bibitem{parrondo2015thermodynamics}
J.~M.~R. Parrondo, J.~M. Horowitz, and T.~Sagawa.
\newblock Thermodynamics of information.
\newblock {\em Nat. Phys.}, 11(2):131--139, 2015.

\bibitem{CB}
S.~Still.
\newblock {Thermodynamic Cost and Benefit of Memory}.
\newblock {\em Phys. Rev. Lett.}, 124:050601, Feb 2020.

\bibitem{IB}
N.~Tishby, F.~Pereira, and W.~Bialek.
\newblock {The information bottleneck method}.
\newblock In B.~Hajek and R.~S. Sreenivas, editors, {\em Proc. 37th Annual
  Allerton Conference}, pages 368--377. University of Illinois, 1999.
\newblock {arXiv:0004057}.

\bibitem{mandal2012work}
D.~Mandal and C.~Jarzynski.
\newblock {Work and information processing in a solvable model of Maxwell's
  demon}.
\newblock {\em Proc. Natl. Acad. Sci. U.S.A.}, 109(29):11641--11645, 2012.

\bibitem{strasberg2017quantum}
P.~Strasberg, G.~Schaller, T.~Brandes, and M.~Esposito.
\newblock Quantum and information thermodynamics: a unifying framework based on
  repeated interactions.
\newblock {\em Phys. Rev. X}, 7(2):021003, 2017.

\bibitem{stopnitzky2019physical}
E.~Stopnitzky, S.~Still, T.~E. Ouldridge, and L.~Altenberg.
\newblock Physical limitations of work extraction from temporal correlations.
\newblock {\em Phys. Rev. E}, 99(4):042115, 2019.

\bibitem{Still-IAL09}
S.~Still.
\newblock Information-theoretic approach to interactive learning.
\newblock {\em EPL}, 85:28005, 2009.

\bibitem{grimsmo2013quantum}
A.~L. Grimsmo.
\newblock Quantum correlations in predictive processes.
\newblock {\em Phys. Rev. A}, 87(6):060302, 2013.

\bibitem{Still-IBPI-2014}
S.~Still.
\newblock Information bottleneck approach to predictive inference.
\newblock {\em Entropy}, 16:968--989, 2014.

\bibitem{StillCruEl10}
S.~Still, J.~P. Crutchfield, and C.~J. Ellison.
\newblock Optimal causal inference: Estimating stored information and
  approximating causal architecture.
\newblock {\em Chaos}, 20:037111, 2010.

\bibitem{wiskott2003slow}
L.~Wiskott.
\newblock {Slow feature analysis: A theoretical analysis of optimal free
  responses}.
\newblock {\em Neural Computation}, 15(9):2147--2177, 2003.

\bibitem{creutzig2009past}
F.~Creutzig, A.~Globerson, and N.~Tishby.
\newblock Past-future information bottleneck in dynamical systems.
\newblock {\em Phys. Rev. E}, 79(4):041925, 2009.

\bibitem{tishby2015deep}
N.~Tishby and N.~Zaslavsky.
\newblock Deep learning and the information bottleneck principle.
\newblock In {\em 2015 IEEE Information Theory Workshop (ITW)}, pages 1--5.
  IEEE, 2015.

\bibitem{shwartz2017opening}
R.~Shwartz-Ziv and N.~Tishby.
\newblock Opening the black box of deep neural networks via information.
\newblock {\em arXiv:1703.00810}, 2017.

\bibitem{bennett1982thermodynamics}
C.~H. Bennett.
\newblock The thermodynamics of computation---a review.
\newblock {\em Int. J. Theor. Phys.}, 21(12):905--940, Dec 1982.

\bibitem{sagawa2009minimal}
T.~Sagawa and M.~Ueda.
\newblock {Minimal energy cost for thermodynamic information processing:
  measurement and information erasure}.
\newblock {\em Phys. Rev. Lett.}, 102(25):250602, 2009.

\bibitem{fahn1996maxwell}
P.~N. Fahn.
\newblock Maxwell's demon and the entropy cost of information.
\newblock {\em Foundations of Physics}, 26(1):71--93, 1996.

\bibitem{ouldridge2018power}
T.~E. Ouldridge, R.~A. Brittain, and P.~R.~t. Wolde.
\newblock The power of being explicit: demystifying work, heat, and free energy
  in the physics of computation.
\newblock {\em arXiv:1812.09572}, 2018.

\bibitem{barkeshli2005dissipationless}
M.~M. Barkeshli.
\newblock Dissipationless information erasure and landauer's principle.
\newblock {\em arXiv:cond-mat/0504323}, 2005.

\bibitem{zurek1989algorithmic}
W.~H. Zurek.
\newblock Algorithmic randomness and physical entropy.
\newblock {\em Phys. Rev. A}, 40(8):4731, 1989.

\bibitem{song2019optimal}
J.~Song, S.~Still, R.~D.~H. Rojas, I.~P. Castillo, and M.~Marsili.
\newblock Optimal work extraction and mutual information in a generalized
  szil{\'a}rd engine.
\newblock {\em Phys. Rev. E}, 103(5):052121, 2021.

\bibitem{zurek1986maxwell}
W.~H. Zurek.
\newblock {Maxwell's demon, Szilard's engine and quantum measurements}.
\newblock In {\em Frontiers of nonequilibrium statistical physics}, pages
  151--161. Springer, Boston (MA), 1986.

\bibitem{lloyd1997quantum}
S.~Lloyd.
\newblock {Quantum-mechanical Maxwell's demon}.
\newblock {\em Phys. Rev. A}, 56(5):3374, 1997.

\bibitem{deffner2016quantum}
S.~Deffner, J.~P. Paz, and W.~H. Zurek.
\newblock Quantum work and the thermodynamic cost of quantum measurements.
\newblock {\em Phys. Rev. E}, 94(1):010103(R), 2016.

\bibitem{Note1}
Notation: entropy and information are functionals of probability distributions.
  But the information theory literature \cite {Cover1991} often adopts the
  following shorthand notation, which we use: $H[U] \equiv -\left \langle
  \protect \qopname \relax o{log}\left [ {p(u)}\right ] \right \rangle _{p(u)}$
  for entropy, $H[U|M] \equiv -\left \langle \protect \qopname \relax
  o{log}\left [ {p(u|m)}\right ] \right \rangle _{p(m,u)}$, for conditional
  entropy, and for mutual information $I[M,U] \equiv \left \langle \protect
  \qopname \relax o{log}\left [ {p(m,u) \over p(m)p(u)} \right ] \right \rangle
  _{p(m,u)} \label {MI}$. Brackets, $\langle \cdot \rangle _p$ denote averages
  over the probability distribution $p$.

\bibitem{sagawa2012thermodynamics}
T.~Sagawa.
\newblock {Thermodynamics of Information Processing in Small Systems}.
\newblock {\em Prog. Theor. Phys.}, 127(1):1--56, 01 2012.

\bibitem{Shannon48}
C.~E. Shannon.
\newblock A mathematical theory of communication.
\newblock {\em Bell Syst. Tech. J.}, 27:379--423, 623--656, 1948.

\bibitem{Berger71}
T.~Berger.
\newblock {\em Rate distortion theory: A mathematical basis for data
  compression}.
\newblock Prentice-Hall, 1971.

\bibitem{StillCru07}
S.~Still and J.~P. Crutchfield.
\newblock Structure or noise?
\newblock {\em arXiv:0708.0654}, 2007.

\bibitem{blahut}
R.~Blahut.
\newblock Computation of channel capacity and rate-distortion functions.
\newblock {\em IEEE transactions on Information Theory}, 18(4):460--473, 1972.

\bibitem{arimoto}
S.~Arimoto.
\newblock An algorithm for computing the capacity of arbitrary discrete
  memoryless channels.
\newblock {\em IEEE Transactions on Information Theory}, 18(1):14--20, 1972.

\bibitem{rose1998deterministic}
K.~Rose.
\newblock Deterministic annealing for clustering, compression, classification,
  regression, and related optimization problems.
\newblock {\em Proceedings of the IEEE}, 86(11):2210--2239, 1998.

\bibitem{wu2020phase}
T.~Wu and I.~Fischer.
\newblock Phase transitions for the information bottleneck in representation
  learning.
\newblock {\em arXiv:2001.01878}, 2020.

\bibitem{Note2}
See supplementary material at [\protect \url
  {https://stacks.iop.org/NJP/24/073031/mmedia}] for a movie of the
  visualization of the physical codebook displayed in figure \ref
  {pistons-codebook}, to higher precision in $x$ and $\tau $.

\bibitem{proesmans2015efficiency}
K.~Proesmans, C.~Driesen, B.~Cleuren, and C.~Van~den Broeck.
\newblock Efficiency of single-particle engines.
\newblock {\em Phys. Rev. E}, 92(3):032105, 2015.

\bibitem{curzon1975efficiency}
F.~L. Curzon and B.~Ahlborn.
\newblock Efficiency of a carnot engine at maximum power output.
\newblock {\em American Journal of Physics}, 43(1):22--24, 1975.

\bibitem{van2005thermodynamic}
C.~Van~den Broeck.
\newblock Thermodynamic efficiency at maximum power.
\newblock {\em Phys. Rev. Lett.}, 95(19):190602, 2005.

\bibitem{schmiedl2007efficiency}
T.~Schmiedl and U.~Seifert.
\newblock Efficiency at maximum power: An analytically solvable model for
  stochastic heat engines.
\newblock {\em EPL (Europhysics Letters)}, 81(2):20003, 2007.

\bibitem{esposito2009universality}
M.~Esposito, K.~Lindenberg, and C.~Van~den Broeck.
\newblock Universality of efficiency at maximum power.
\newblock {\em Phys. Rev. Lett.}, 102(13):130602, 2009.

\bibitem{esposito2010efficiency}
M.~Esposito, R.~Kawai, K.~Lindenberg, and C.~Van~den Broeck.
\newblock Efficiency at maximum power of low-dissipation carnot engines.
\newblock {\em Physical review letters}, 105(15):150603, 2010.

\bibitem{allahverdyan2013carnot}
A.~E. Allahverdyan, K.~V. Hovhannisyan, A.~V. Melkikh, and S.~G. Gevorkian.
\newblock Carnot cycle at finite power: Attainability of maximal efficiency.
\newblock {\em Phys. Rev. Lett.}, 111(5):050601, 2013.

\bibitem{PhysRevLett.117.190601}
N.~Shiraishi, K.~Saito, and H.~Tasaki.
\newblock Universal trade-off relation between power and efficiency for heat
  engines.
\newblock {\em Phys. Rev. Lett.}, 117:190601, Oct 2016.

\bibitem{PhysRevLett.120.190602}
P.~Pietzonka and U.~Seifert.
\newblock Universal trade-off between power, efficiency, and constancy in
  steady-state heat engines.
\newblock {\em Phys. Rev. Lett.}, 120:190602, May 2018.

\bibitem{deffner2018efficiency}
S.~Deffner.
\newblock Efficiency of harmonic quantum otto engines at maximal power.
\newblock {\em Entropy}, 20(11):875, 2018.

\bibitem{hong2020quantum}
Y.~Hong, Y.~Xiao, J.~He, and J.~Wang.
\newblock Quantum otto engine working with interacting spin systems: Finite
  power performance in stochastic thermodynamics.
\newblock {\em Phys. Rev. E}, 102(2):022143, 2020.

\bibitem{sagawa2008second}
T.~Sagawa and M.~Ueda.
\newblock Second law of thermodynamics with discrete quantum feedback control.
\newblock {\em Phys. Rev. Lett.}, 100(8):080403, 2008.

\bibitem{sagawa2010generalized}
T.~Sagawa and M.~Ueda.
\newblock Generalized jarzynski equality under nonequilibrium feedback control.
\newblock {\em Phys. Rev. Lett.}, 104(9):090602, 2010.

\bibitem{sagawa2012nonequilibrium}
T.~Sagawa and M.~Ueda.
\newblock Nonequilibrium thermodynamics of feedback control.
\newblock {\em Phys. Rev. E}, 85(2):021104, 2012.

\bibitem{esposito2012stochastic}
M.~Esposito and G.~Schaller.
\newblock Stochastic thermodynamics for “maxwell demon” feedbacks.
\newblock {\em EPL (Europhysics Letters)}, 99(3):30003, 2012.

\bibitem{abreu2012thermodynamics}
D.~Abreu and U.~Seifert.
\newblock Thermodynamics of genuine nonequilibrium states under feedback
  control.
\newblock {\em Phys. Rev. Lett.}, 108(3):030601, 2012.

\bibitem{barato2014unifying}
A.C. Barato and U.~Seifert.
\newblock Unifying three perspectives on information processing in stochastic
  thermodynamics.
\newblock {\em Phys. Rev. Lett.}, 112(9):090601, 2014.

\bibitem{debiossac2020thermodynamics}
M.~Debiossac, D.~Grass, J.~J. Alonso, E.~Lutz, and N.~Kiesel.
\newblock Thermodynamics of continuous non-markovian feedback control.
\newblock {\em Nat. Commun.}, 11(1):1--6, 2020.

\bibitem{cao2009thermodynamics}
F.~J. Cao and M.~Feito.
\newblock Thermodynamics of feedback controlled systems.
\newblock {\em Phys. Rev. E}, 79(4):041118, 2009.

\bibitem{strasberg2019stochastic}
P.~Strasberg and A.~Winter.
\newblock Stochastic thermodynamics with arbitrary interventions.
\newblock {\em Phys. Rev. E}, 100(2):022135, 2019.

\bibitem{horowitz2014second}
J.~M. Horowitz and H.~Sandberg.
\newblock Second-law-like inequalities with information and their
  interpretations.
\newblock {\em New J. Phys.}, 16(12):125007, 2014.

\bibitem{bechhoefer2015hidden}
J.~Bechhoefer.
\newblock Hidden markov models for stochastic thermodynamics.
\newblock {\em New J. Phys.}, 17(7):075003, 2015.

\bibitem{crooks2019marginal}
G.~E. Crooks and S.~Still.
\newblock Marginal and conditional second laws of thermodynamics.
\newblock {\em EPL (Europhysics Letters)}, 125(4):40005, 2019.

\bibitem{Note3}
In this exploration, we broke off the numerical optimization at the point where
  we observed a transition to four memory states, because the trends are
  clearly visible at that point. We thus show only solutions up to three memory
  states. We calculated parametrically optimized deterministic partitions at
  $\tau $ intervals of 1, which is coarser than the spacing used to compute
  optimal memories.

\bibitem{Cover1991}
T.~M. Cover and J.~A. Thomas.
\newblock {\em Elements of information theory}.
\newblock Wiley, New York, 1991.

\end{thebibliography}
\end{document}